\documentclass[
 aps,
 prb,
 preprint,
 superscriptaddress,
 amsmath,amssymb,
 longbibliography,
]{revtex4-2}

\usepackage{graphicx}
\usepackage{dcolumn}
\usepackage{color}
\usepackage{hyperref}
\usepackage{lineno}
\usepackage{gensymb}

\begin{document}

\title{Interaction of water with nitrogen-doped graphene}

\author{Azim Fitri Zainul Abidin}
\affiliation{
Department of Precision Engineering,
Graduate School of Engineering,
Osaka University,
2-1 Yamadaoka, Suita, Osaka 565-0871, Japan
}


\author{Ikutaro Hamada}%
\email{ihamada@prec.eng.osaka-u.ac.jp}
\affiliation{
Department of Precision Engineering,
Graduate School of Engineering,
Osaka University,
2-1 Yamadaoka, Suita, Osaka 565-0871, Japan
}

\date{\today}

\begin{abstract}
We have studied the interaction of water and graphene doped with nitrogen in different configurations, namely, graphitic and pyridinic nitrogen, by means of the van der Waals density functional.
We found that the local nitrogen configuration plays a key role in determining the stable water configuration, while the dispersion force is responsible for the water adsorption.
With the graphitic nitrogen, water prefers to orient with its oxygen toward the surface, whereas for the pyridinic nitrogen it prefers to orient with its hydrogens toward the surface, because nitrogen is positively and negatively charged for the former and the latter, respectively.
Our results have great implications for the modeling of the interface between water and nitrogen-doped graphitic systems.
\end{abstract}

\maketitle

\clearpage

\section{Introduction}
\label{sec:introduction}
%
%
Graphene has been one of the most attractive materials since its discovery, due to its unique structural, electronic, and mechanical properties \cite{neto2009electronic,novoselov2004electric,Novoselov2005,georgakilas2016noncovalent,geim2010rise,blonski2017doping}.
It has also been found to be useful and effective in a broad class of applications, including energy conversion and storage and sensing \cite{georgakilas2012functionalization,georgakilas2016noncovalent}.
Doping graphene with foreign elements has been investigated to tailor its electronic properties, for instance, opening up its zero band gap to provide new possibilities for more applications \cite{de2018non}.
Nitrogen (N) is one of the most frequently studied dopants for modifying the electronic, transport, and magnetic properties of graphene.
In the catalytic research field, N doping of graphene has been utilized to improve the catalytic activities of electrochemical reactions, in particular, the oxygen reduction reaction (ORR) for the fuel cell application \cite{guo2016active,liang2014hierarchically,fitri2017cobalt,abidin2018effect,abidin2019nitrogen}.
It was reported that the intrinsic physico-chemical properties and catalytic reactivity of N-doped graphene are governed by the N concentration as well as its local structures (e.g., graphitic- and pyridinic-N configurations) and environment \cite{lin2015structural,deng2016catalysis,jiao2016activity,wu2016metal,wu2016incorporation,marshall2020pyridinic}.
Furthermore, it was reported that the phenomena at the phase boundary of the graphene layer are significantly affected by the N doping.
However, the role of the N doping of graphene in catalytic reactions has not been fully understood.
%
%
%

%
To investigate the N-doped graphene as a catalyst for an electrochemical reaction such as ORR, it is essential to understand its interaction with water, as it plays the central role in the electrochemistry.
The water-graphene interaction has been the subject of intense research.
Yet, the interaction is still unclear as it is highly dependent on surface properties.
Most of these studies concluded that the graphene is hydrophobic and its wetting properties are similar to those of graphite \cite{li2013effect,kim2011chemical,shin2010surface}, while some studies suggested that a clean monolayer graphene is considered hydrophilic.
In the latter case, it was reported that the graphene becomes hydrophobic as the number of graphene layers increases \cite{kozbial2014understanding,munz2015thickness}.
It was also reported that water layers are formed on graphene \cite{ho2013polarizability,akaishi2017formation} as well as carbon nanotubes \cite{ho2013polarizability}.
Several recent studies have also reported that doping graphene by electrical or chemical means can lift or enhance its wetting properties \cite{hong2016mechanism,shih2012breakdown,rafiee2010superhydrophobic}.
Furthermore, it has been suggested that hydrophobicity/hydrophilicity in the vicinity of the active site of N-doped carbon catalysts plays a crucial role in the catalytic activity \cite{Singh2019,Takeyasu2021}.
Thus, it is important to understand the interaction between water and N-doped graphene in addition to the pristine one, to understand the role of the N dopant in the important electrochemical reactions.
The density functional theory (DFT) calculations have been routinely used to investigate the interaction of water with graphitic materials.
DFT simulations allow us to study the microscopic properties of water confined in the graphene materials, which are difficult to access experimentally, and complement the experimental measurements.
However, it is well known that the conventional semilocal DFTs cannot capture the long-range dispersion interaction, which is crucial to describe the water/graphene interfaces.
Thus, a highly accurate DFT exchange-correlation functional is required to precisely calculate the interaction between graphene and water.
Recent development of semiempirical dispersion correction methods \cite{Hermann2017} and the nonlocal van der Waals density functional (vdW-DF) \cite{Berland2015} makes it possible to describe the weak dispersion interaction very accurately within DFT.
Furthermore, thanks to the methodological and algorithmic development, it has become possible to perform highly accurate wave function based electronic structure calculations, allowing the rigorous benchmark of the approximate exchange-correlation functionals.
Quite recently, Brandenburg \textit{et al}. \cite{brandenburg2019interaction} performed an extensive benchmark of dispersion corrected exchange-correlation functionals against the quantum Monte Carlo results for a set of systems containing water and carbon-based materials, and suggested several accurate exchange-correlation functionals for water-carbon systems.

%
%
%
%
%

In this work, we investigate in detail the nature of the interaction between the water molecule and N-doped graphene by means of vdW-DF.
We consider different N configurations, namely, graphitic- and pyridinic-N configurations.
%
%
%
%
%
%
%
Through the comprehensive electronic structure analysis, we clarify the factor governing the interaction between water and N-doped graphene.
%
%
\section{Computational Details}
\label{sec:methods}
\begin{figure}[h]
   \includegraphics[width=0.95\columnwidth]{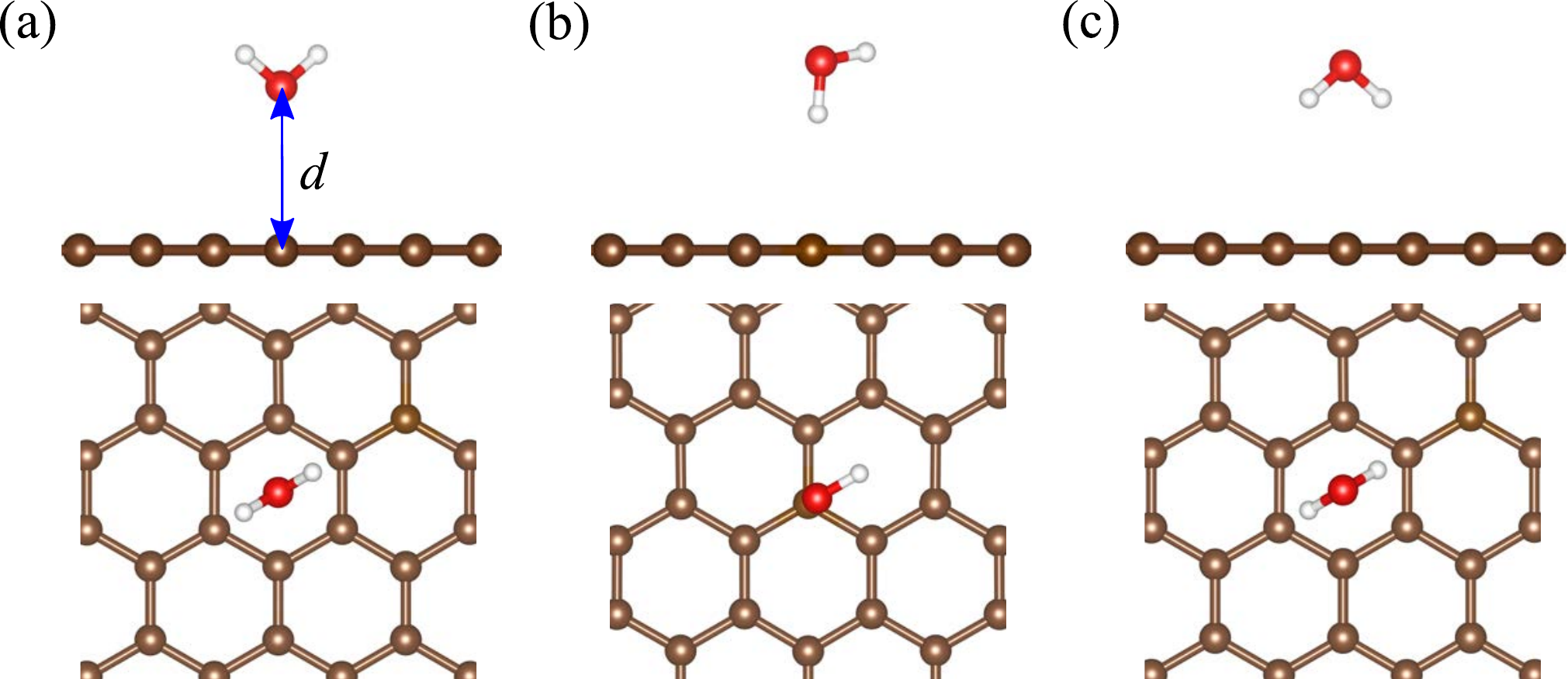}
    \caption{\label{fig:h2o_configuration}Side (upper panel) and top (lower panel) views of the water adsorption configuration on pristine graphene for  (a) 0-leg, (b) 1-leg, and (c) 2-leg configurations. 
    In the 1-leg configuration, water molecule is inclined by 4.51$^{\circ}$ with respect to the surface normal.
    H, C, and O atoms are represented by white, brown, and red spheres, respectively.}
\end{figure}

All the calculations were performed using the projector augmented wave (PAW) \cite{blochl1994projector} method as implemented in the \textsc{Quantum-ESPRESSO} code \cite{giannozzi2009quantum}.
The PAW potentials were adopted from the \textsc{PSLIBRARY} \cite{dal2014pseudopotentials} version 1.0.0.
Wave functions and augmentation charge density were expanded in terms of a plane-wave basis set with the kinetic energy cutoffs of 80 and 800 Ry, respectively.
For the exchange correlation, the rev-vdW-DF2 \cite{hamada2014van} functional, a variant of the van der Waals density functional (vdW-DF) \cite{Dion2004,Thonhauser2007,Berland2015}, as well as the Perdew-Ernzerhof-Burke (PBE) \cite{perdew1996generalized} generalized gradient approximation functional, were used.
The rev-vdW-DF2 is shown to be reasonably accurate in describing the interactions between water and graphitic materials \cite{brandenburg2019interaction}.
The Marzari-Vanderbilt cold smearing \cite{marzari1999thermal} with a smearing width of 0.02 Ry was used to treat the Fermi surface.
Spin polarization was taken into account for the case of water adsorption on graphene doped with graphitic N and pyridinic N.
We confirmed the validity of our computational setup by performing a convergence study with respect to the kinetic energy cutoffs, $\mathbf{k}$-point sampling, with different smearing functions and smearing widths, using the water/graphene structures contained in the \textsc{WaC18} data set \cite{brandenburg2019interaction}.
%
We also confirmed that with the computational setup described above, we reproduce the interaction energies reported in Ref.~\onlinecite{brandenburg2019interaction}.
See the Supplemental Material \cite{SM} for the results of the benchmark calculations.
%
In our production calculations, the lattice constant of graphene was optimized using the PBE functional and 16 $\times$ 16 $\mathbf{k}$-point mesh for Brillouin zone sampling, to obtain 2.467~\AA.
For the water adsorption, a $(5\times5)$ hexagonal supercell was constructed using the PBE-optimized lattice constant
and the structures of adsorbed water were taken from the {\sc WaC18} data set
(Fig.~\ref{fig:h2o_configuration}).
In the rev-vdW-DF2 calculation, we used the graphene supercell constructed using the PBE-optimized lattice constant.
We confirmed the effect of the lattice constant used is negligible, by performing the calculation of water on pristine graphene optimized using rev-vdW-DF2 (see Supplemental Material \cite{SM}).
%
%
%
To eliminated the spurious electrostatic interaction with the periodic images, the effective screening medium method \cite{otani2006first,hamada2009green} was employed, along with a vacuum spacing of 20~\AA.
%
%
%
%
%
%
Geometries of N-doped graphene and water molecules in the gas phase were fully optimized until residual forces on the constituent atoms became smaller than $10^{-4}$ Ry/bohr ($2.57 \times 10^{-3}$ eV/\AA).
We calculated the interaction energy as a function of distance between water and graphene structure defined by
\begin{equation}
E_{\mathrm{int}} = E_{\mathrm{W/G}} - E_{\mathrm{W}} - E_{\mathrm{G}},
\label{eqn:eint}
\end{equation}
where
$E_{\mathrm{W/G}}$, $E_{\mathrm{W}}$, and $E_{\mathrm{G}}$, are the total energies of the combined system, gas-phase water molecule, and isolated graphene structure, respectively.
The distance between the water molecule and graphene structure is defined by the difference between $z$ coordinates of the oxygen (O) atom in water and graphene structure.
In the calculations of the interaction energy curve, the geometries of the water molecule and graphene were fixed to those obtained in their isolation, and the equilibrium geometry (distance) was determined by the cubic spline interpolation.
We also performed full structural optimization for the most stable adsorption configurations and confirmed that the effect of the structural relaxation is minor and our conclusion does not change (see Supplemental Material \cite{SM}).
%
%
Only the $\Gamma$ point was used to sample the surface Brillouin zone in the calculations of the interaction energy curves.

\section{Result and Discussion}
\label{sec:results}
\subsection{Pristine graphene}
\label{sec:gr}
\begin{figure}[ht]
    \includegraphics[width=0.7\columnwidth]{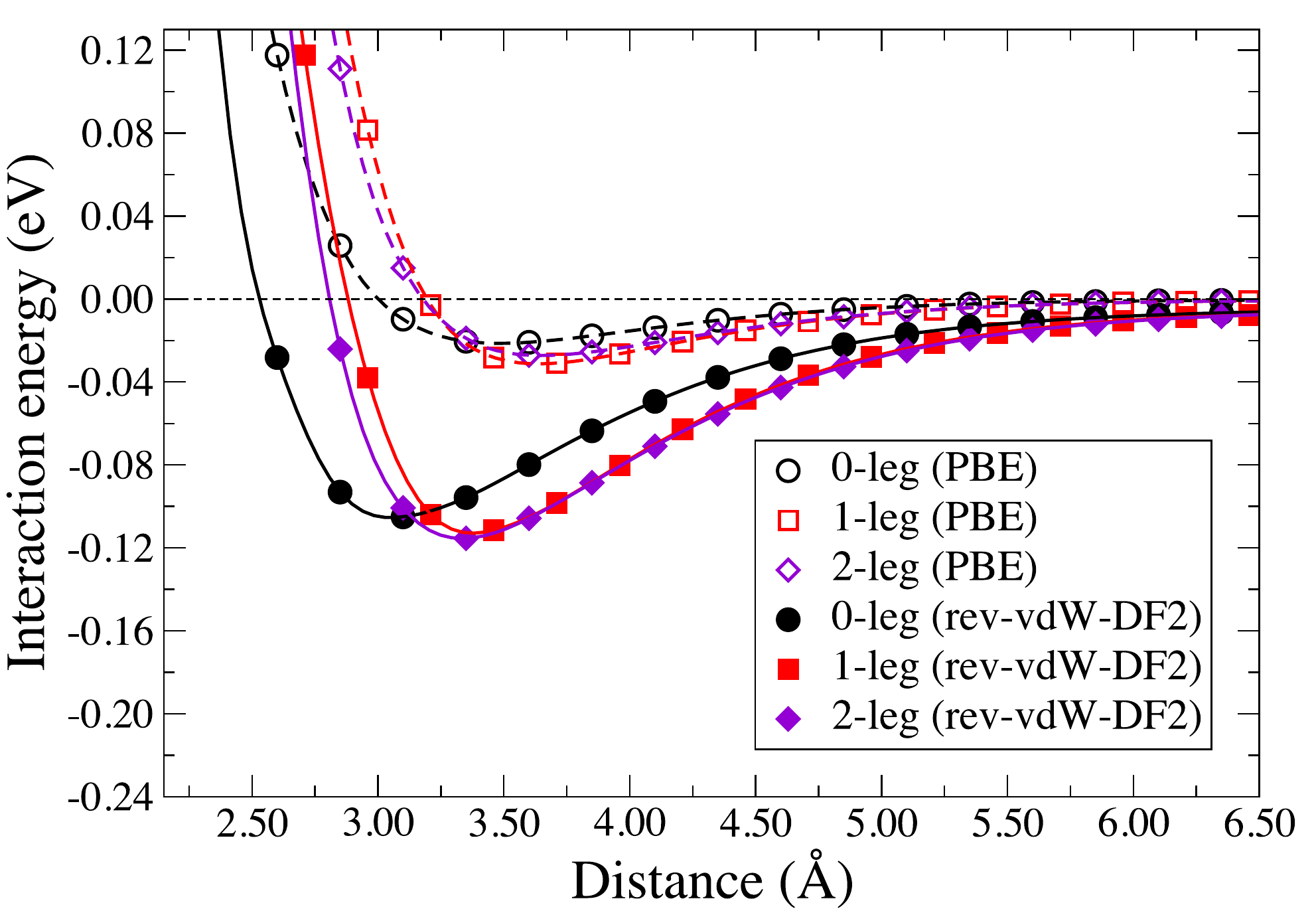}
    \caption{\label{fig:eint_graphene}Interaction energy of water with pristine graphene as a function water-graphene distance.}
 \end{figure}

We first calculated the interaction energy as a function of distance between the water molecule and pristine graphene for 0-leg, 1-leg and 2-leg configurations with PBE and rev-vdW-DF2 functionals (Fig.~\ref{fig:eint_graphene}).
As for the adsorption site, we considered bridge, hollow, and on-top sites, and report the results for the most stable ones (hollow, on-top, and hollow sites, for 0-leg, 1-leg, and 2-leg configurations, respectively), and results for other adsorption sites are reported in the Supplemental Material \cite{SM}.
%
%
We also considered a configuration in which the molecular axis is parallel to the surface, to find it less stable (see Supplemental Material \cite{SM}).
The equilibrium interaction energies ($E_0$) and distances ($d_0$) calculated with PBE and rev-vdW-DF2, along with those obtained using the diffusion Monte Carlo (DMC) method and adiabatic-connection fluctuation-dissipation theorem within the random phase approximation (RPA), are summarized in Table~\ref{tab:gra-dmc}.
%
%
The present PBE results are consistent with the previous work \cite{hamada2012adsorption}.
Overall, the adsorption energies obtained using rev-vdW-DF2 are much larger than those obtained using PBE, and larger than those obtained using vdW-DF2$^{\mathrm{C09}_x}$ (vdW-DF2-C09) \cite{hamada2012adsorption}.
However, the rev-vdW-DF2 values are much closer to DMC and RPA values than those obtained using other vdW-DF functionals (see Refs.~\onlinecite{hamada2012adsorption,brandenburg2019interaction}).
%
We also assessed the accuracy of the recent vdW-DF functionals including vdW-DF-cx \cite{Berland2014}, vdW-DF3 \cite{Chakraborty2020}, and hybrid versions of vdW-DF-cx (vdW-DF-cx0) \cite{Berland2017,Jiao2018,Hyldgaard2020} and rev-vdW-DF2 (rev-vdW-DF2-0), using the water/graphene structures in {\sc WAC18} \cite{brandenburg2019interaction}.
The mean absolute deviations of the interaction energies with respect to the reference DMC results are 18, 21, 41, 30, 32, 16, 19 meV for rev-vdW-DF2, vdW-DF2-C09, vdW-DF-cx, vdW-DF3-opt1, vdW-DF3-opt2, vdW-DF-cx0, and rev-vdW-DF2-0, respectively. See Supplemental Material \cite{SM} for details.
%
The equilibrium distances obtained using rev-vdW-DF2 are also in good agreement with the DMC and RPA ones.
Given that the results for pristine graphene are reasonably accurate, we decided to use rev-vdW-DF2 as a vdW-inclusive exchange-correlation functional to further investigate the interaction of water with N-doped graphene.
In the following, PBE results are also shown for comparison.
 \begin{table*}[ht]
\caption{\label{tab:gra-dmc} Equilibrium interaction energy ($E_{0}$) and distance ($d_{0}$) of water on pristine graphene obtained using PBE and rev-vdW-DF2 functionals, along with DMC and RPA values.}
\begin{ruledtabular}
\begin{tabular}{ccccccccc}
              &\multicolumn{2}{c}{PBE} & \multicolumn{2}{c}{rev-vdW-DF2} & \multicolumn{2}{c}{DMC\footnotemark[1]} & \multicolumn{2}{c}{RPA\footnotemark[1]}\\
              \cline{2-3} \cline{4-5} \cline{6-7} \cline{8-9}
Configuration & $E_{0}$ & $d_0$ & $E_{0}$ & $d_0$ & $E_{0}$ & $d_0$ & $E_{0}$ & $d_0$\\
              & (meV) & (\AA) & (meV) & (\AA)  & (meV) & (\AA)  & (meV) & (\AA) \\
\hline
0-leg &$-$22 & 3.513  &$-$106 &3.049 &$-$90 $\pm$ 6 &3.10 $\pm$ 3 &$-$90 $\pm$ 2 &3.05 $\pm$ 1\\
1-leg &$-$32 & 3.672  &$-$113 &3.401 &$-$92 $\pm$ 6 &3.46 $\pm$ 3 &$-$87 $\pm$ 1 &3.45 $\pm$ 1\\
2-leg &$-$27 & 3.608  &$-$116 &3.321 &$-$99 $\pm$ 6 &3.37 $\pm$ 2 &$-$98 $\pm$ 1 &3.33 $\pm$ 1\\
\end{tabular}
\end{ruledtabular}
\footnotetext[1]{Ref.~\onlinecite{Brandenburg2019}}
\end{table*}
%
%

%
\subsection{Nitrogen doped graphene}
\label{sec:gr-n}
\begin{figure}[ht]
    \includegraphics[width=0.95\columnwidth]{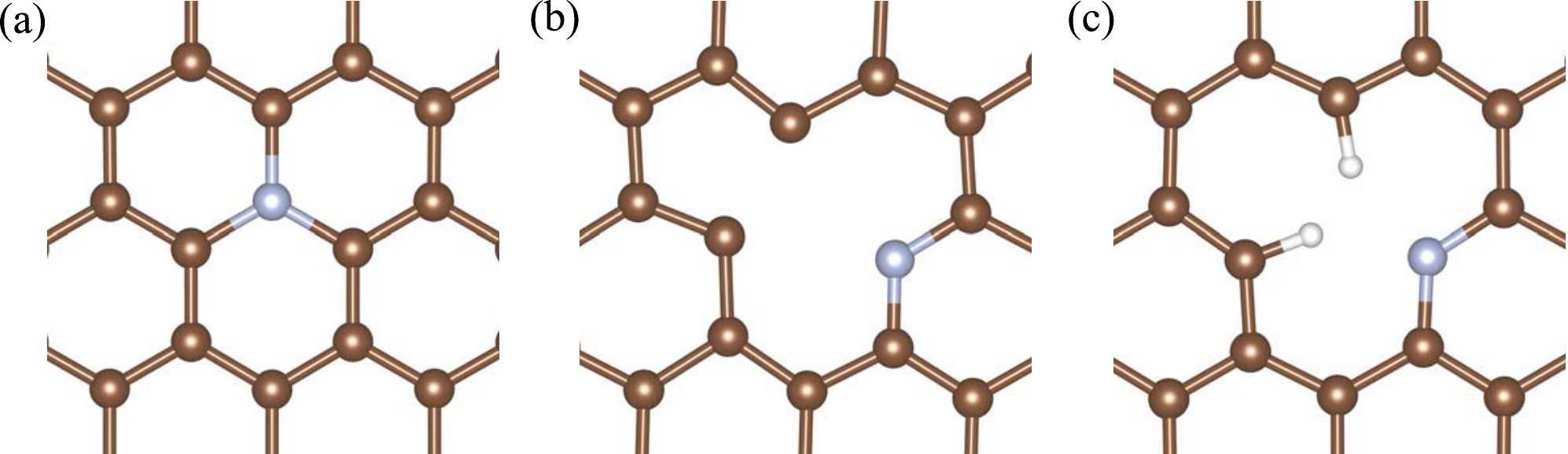}
    \caption{\label{fig:cn_structure}Structures of graphene doped with (a) graphitic N, (b) pyridinic N and dangling C, and (c) pyridinic N and H-terminated C. White, brown, and silver spheres represent H, C, and N atoms, respectively.}
\end{figure}

We then investigated the water adsorption on N-doped graphene.
We chose the N-doped graphene structures of the current most interest, namely, those doped with graphitic N [gr-N; Fig.~\ref{fig:cn_structure}(b)], with pyridinic N with dangling (non-H-terminated) carbon (C) atoms [pyri-N (no H); Fig.~\ref{fig:cn_structure}(c)], and pyridinic N with H-terminated C atoms [pyri-N; Fig.~\ref{fig:cn_structure}(d)].
The gr-N configuration was constructed by replacing one C atom with a N atom in graphene.
For the variants of the pyri-N structure, monovacancy was initially created in graphene and then one of the C atoms next to the vacancy site was replaced with N, to create the pyri-N (no H) configuration.
Subsequently, two C atoms with dangling bonds were terminated by hydrogen (H) atoms, to form the pyri-N configuration.
As in the case of pristine graphene, water adsorptions in 0-leg [Fig.~\ref{fig:h2o_configuration}(a)], 1-leg [Fig.~\ref{fig:h2o_configuration}(b)], 2-leg configurations [Fig.~\ref{fig:h2o_configuration}(c)] were investigated.
\subsubsection{Graphitic nitrogen}
\label{sec:gr-gr-n}

\begin{figure}
    \includegraphics[width=0.95\columnwidth]{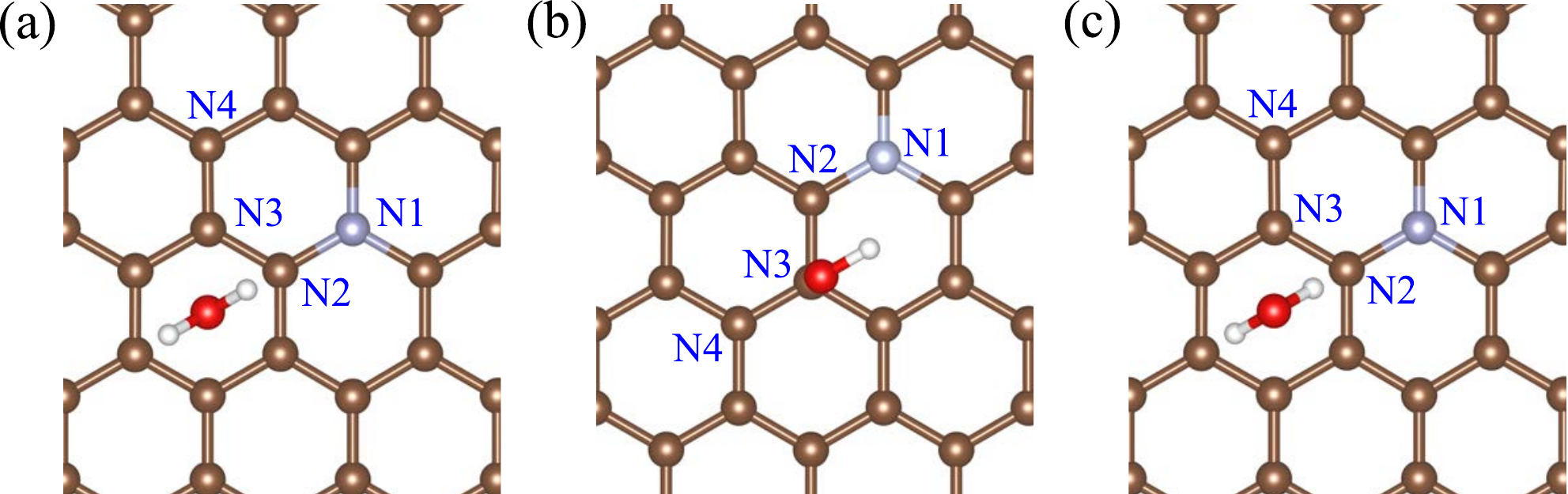}
    \caption{\label{fig:grn-water}Water adsorption configuration on graphene doped with graphitic N with difference N positions (N1-N4) for (a) 0-leg, (b) 1-leg, and (c) 2-leg configurations.}
 \end{figure}

\begin{figure}
    \includegraphics[width=1.0\columnwidth]{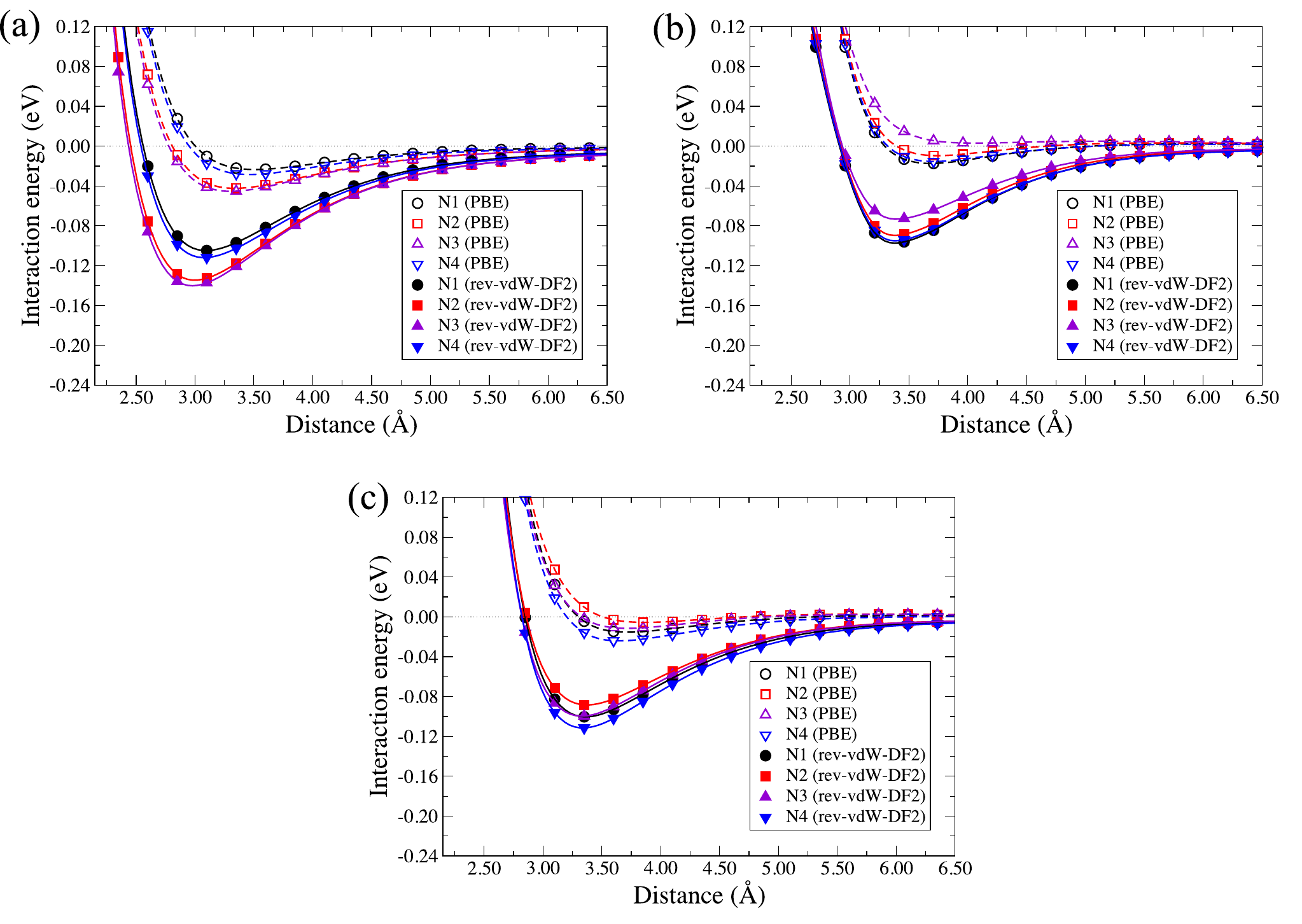}
    \caption{\label{fig:grn-int}Interaction energy of water on graphene doped with graphitic N as a function of water-graphene distance with a different N positions for (a) 0-leg, (b) 1-leg, and (c) 2-leg configurations.}
 \end{figure}

We start with the interaction between water and graphene doped with gr-N.
Different relative N positions were considered, namely, N1, N2, N3, and N4 (Fig.~\ref{fig:grn-water}), in addition to the water configurations.
The interaction energies calculated as functions of water-graphene distance using PBE and rev-vdW-DF2 functionals are shown in Fig.~\ref{fig:grn-int}.
The equilibrium interaction energies and distances are summarized in Table. ~\ref{tab:grn-int}.
As in the case of pristine graphene, rev-vdW-DF2 consistently predicts much more attractive interaction between water and N-doped graphene than PBE, and the nonlocal correlation, i.e., the dispersion interaction, is responsible for the water adsorption.
See Supplemental Material \cite{SM} for decomposition of the interaction energy.
It should be noted that PBE predicts attractive interaction between graphene and water, although PBE is not able to describe the dispersion interaction accurately.
This is because of the spurious binding due to the exchange functional used in the PBE functional as discussed in Refs.~\onlinecite{Wu2001,Murray2009,Hamada2010}.
The calculated interaction energies indicate that the doped N induces different effects on the stability of water on graphene, depending on the configuration and the N position, and the former has a larger effect than the latter.
We found that on graphene doped with gr-N, the 0-leg configuration is the most stable one with the favorable positions at N3 as well as N2 (the positions close to the adsorbed water molecule) in the case of the hollow site adsorption.
In a recent publication, Pham {\it et al.} \cite{Pham2021} reported water adsorption on top of gr-N in graphene.
We also examined water adsorption on top of gr-N and found the configuration is slightly more stable with $E_{\mathrm{ads}}= - 152 $ meV, agreeing well with that obtained by Pham {\it et al.} \cite{Pham2021} (Supplemental Material \cite{SM}).
Yet, the energy difference is tiny and our conclusion does not change when we consider the hollow site adsorption.
Both PBE and rev-vdW-DF2 predict that the adsorption of water with the O atom pointing toward the surface (0-leg configuration) is strengthened in the vicinity of doped N.
These results imply that the enhancement is mainly attributed to the electrostatic effect, as PBE cannot describe the dispersion interaction properly (but does the electrostatics) and that the doped N is positively charged.
This is further supported by the facts that the interaction of water in the 1-leg and 2-leg configurations with graphene is much weaker.
The interaction is particularly weak for water in the 1-leg configuration interacting directly with the N atom underneath (N3 position).
%
\begin{table*}
\caption{\label{tab:grn-int}Equilibrium interaction energies ($E_{0}$) and distance $d_{0}$ of water on graphene doped with graphitic N with difference N-dopant positions for 0-leg, 1-leg, and 2-leg configurations obtained using the PBE and rev-vdW-DF2 functionals.}
\begin{ruledtabular}
\begin{tabular}{ccccccc}
&\multicolumn{2}{c}{0-leg} &\multicolumn{2}{c}{1-leg} &\multicolumn{2}{c}{2-leg}\\
N position & $E_{0}$ & $d_{0}$ & $E_{0}$ & $d_{0}$ & $E_{0}$ & $d_{0}$ \\
           & (meV)   & (\AA)   & (meV)   & (\AA)   & (meV) & (\AA)     \\ \hline
& \multicolumn{6}{c}{PBE} \\
N1 &$-$24 &3.512 &$-$18  &3.672 &$-$17 &3.720\\
N2 &$-$43 &3.342 &$-$10  &3.709 &$-$6 &3.784\\
N3 &$-$46 &3.347 &   +3  &3.816 &$-$13 &3.736\\
N4 &$-$29 &3.501 &$-$16  &3.688 &$-$25 &3.645\\
& \multicolumn{6}{c}{rev-vdW-DF2}\\
N1 &$-$105 &3.086 &$-$97 &3.353 &$-$101 &3.335\\
N2 &$-$135 &3.022 &$-$91 &3.378 &$-$80 &3.352\\
N3 &$-$141 &2.991 &$-$74 &3.395 &$-$100 &3.337\\
N4 &$-$112 &3.081 &$-$96 &3.368 &$-$112 &3.331\\
\end{tabular}
\end{ruledtabular}
\end{table*}
We calculated the densities of states of the adsorbed systems for pristine and N-doped graphenes, and the wave functions corresponding to the molecular orbitals (MOs) of water molecule in the most stable adsorption configuration.
We found that the hybridization of the MOs with the the wave functions of graphene is negligibly small, and the character of the water MOs is retained upon adsorption, implying that there is no significant bonding between water and substrate states (see Supplemental Material \cite{SM}).

\begin{figure}[h]
   \includegraphics[width=0.8\textwidth]{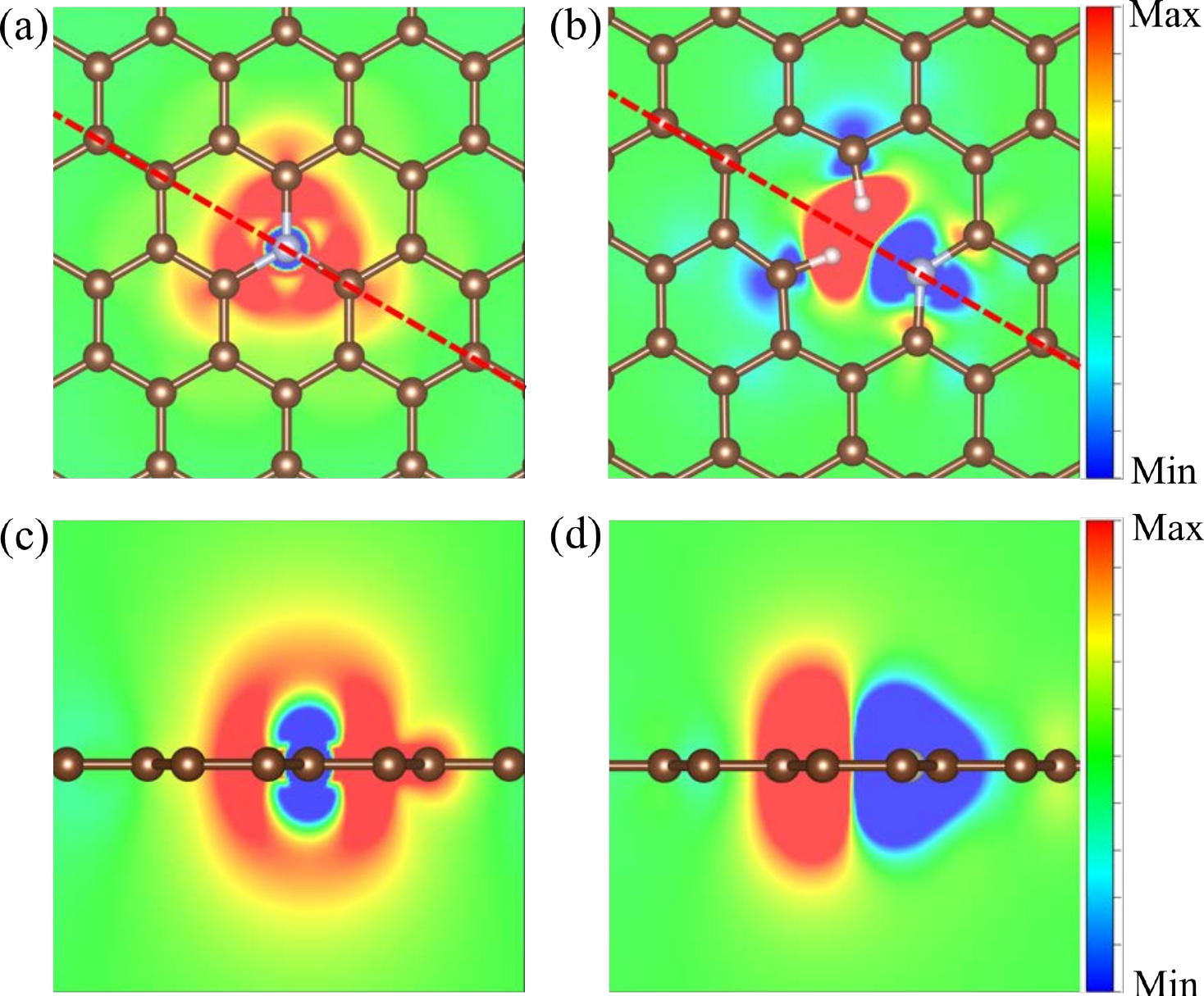}
   \caption{\label{fig:dvhar}Hartree potential difference ($\Delta V_{\mathrm{H}}$) for graphitic N [(a) top and (c) side views] and that for pyridinic N [(b) top and (d) side views]. The maximum (minimum) value of $\Delta V_{\mathrm{H}}$ is 1.1 ($-$0.5) eV.  The red dotted line indicates the lattice plane for the side views.}
 \end{figure}

To gain further insight into the interaction between water and N-doped graphene, we calculated the charge-density difference defined by $\Delta\rho(\textbf{r})=\rho_{\mathrm{W/G}}(\textbf{r})-\Delta\rho_{\mathrm{W}}(\textbf{r})-\Delta\rho_{\mathrm{G}}(\textbf{r})$, where $\Delta\rho_{\mathrm{W/G}}(\textbf{r})$, $\Delta\rho_{\mathrm{W}}(\textbf{r})$, and $\Delta\rho_{\mathrm{G}}(\textbf{r})$ are the charge densities of the total adsorption system, isolated water molecule, and isolated graphene structure, respectively.
Figure ~\ref{fig:drho_graphitic_n} shows the calculated $\Delta \rho$'s for each configurations (see Supplemental Material \cite{SM} for other configurations).
We observed a large variation of the charge polarization, depending on the water configuration.
In the 0-leg configuration, the electronic charge accumulates (negatively charged) in the vicinity of the O atom of water, while the charge is depleted (positively charged) in the vicinity of C and N atoms near the adsorption site.
The polarization is significant and thus the adsorption is stabilized, when the doped N atom is close to the adsorption site.
On the other hand, in the cases of 1-leg and 2-leg configurations, the electronic charge accumulates at the C and/or N sites, to which H atoms of water are directed.
We can see that the polarization of the charge in graphene is small, when the doped N atom is close to the H atoms of water (N3 in the 1-leg configuration and N2 in the 2-leg configuration), indicating that the doped N atom is less polarizable negatively.
Based on the water configuration and charge density difference, we concluded that the gr-N in graphene is positively charged.
%
We note that our Bader charge analysis indicates that gr-N is negatively charged (see Supplemental Material \cite{SM}), but this does not explain the orientation of the water molecule.
 \begin{figure}[ht]
   \includegraphics[width=0.95\textwidth]{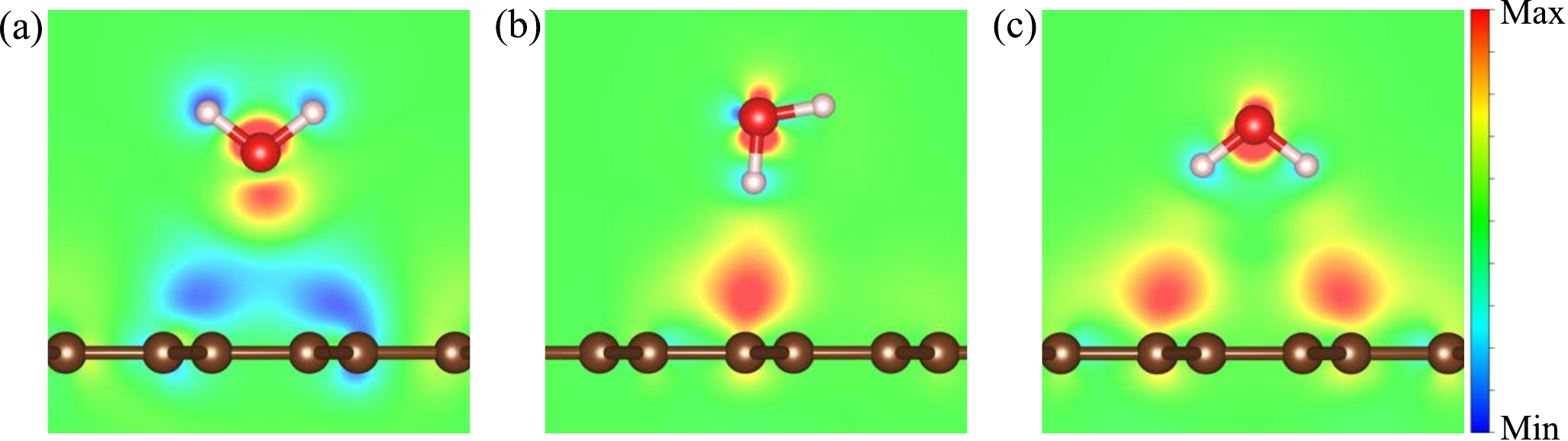}
        \caption{\label{fig:drho_graphitic_n}Charge-density difference ($\Delta\rho$) of water on graphitic N for (a) 0-leg, (b) 1-leg, and (c) 2-leg configurations with the most stable relative N positions (N3, N1, and N4, respectively). $\Delta\rho$ in a plane containing the molecular plane is plotted. The maximum (minimum) value of $\Delta \rho$ is $1.35\times10^{-2}$ ($-1.35\times10^{-2}$) $e$ \AA$^{-3}$.}
 \end{figure}
To further verify the charge state of the gr-N in graphene, we calculated the Hartree potential difference defined by
$\Delta{V}_{\mathrm{H}}(\textbf{r})= V_{\mathrm{H}}^{\mathrm{G}}(\textbf{r}) - V_{\mathrm{H}}^ {\mathrm{GN}}(\textbf{r})$, where $V_{\mathrm{H}}^{\mathrm{G}}(\textbf{r})$ and $V_{\mathrm{H}}^ {\mathrm{GN}}(\textbf{r})$ are the electrostatic (Hartree) potential of pristine and N-doped graphene structures, respectively \cite{mallada2020atomic}.
Figures \ref{fig:dvhar} (a) and
(c)
show two-dimensional plots of $\Delta V_{\mathrm{H}}$ in the planes containing doped N parallel and normal to graphene.
We can see that the potential difference at and around the N nucleus is negative, while it is positive in its vicinity, suggesting that the N and surround C atoms are positively charged.
Note that the valence pseudo-charge-densities were used in the calculation of $\Delta V_{\mathrm{H}}$, and thus the potentials near the nuclei are less accurate and reliable.
The result of our analysis is consistent with that reported in Ref.~\onlinecite{mallada2020atomic}.
%
%
%
%
%

%
\subsubsection{Pyridinic nitrogen}
\label{sec:gr-pyri-n}

\begin{figure}
    \includegraphics[width=0.95\columnwidth]{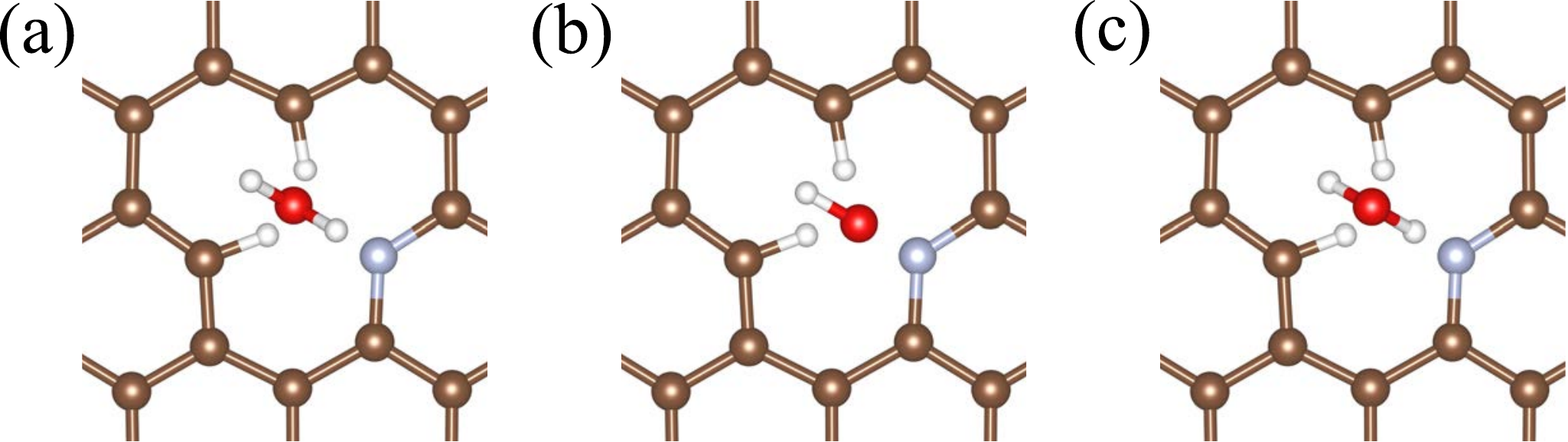}
    \caption{\label{fig:pyri-n}  Water adsorption configurations on graphene doped with pyridinic N and H-terminated C for (a) 0-leg configuration, (b) 1-leg configuration (orientation 1) (1-leg 1), and (c) 2-leg configuration (orientation 1).}
\end{figure}

\begin{figure}
    \includegraphics[width=0.8\columnwidth]{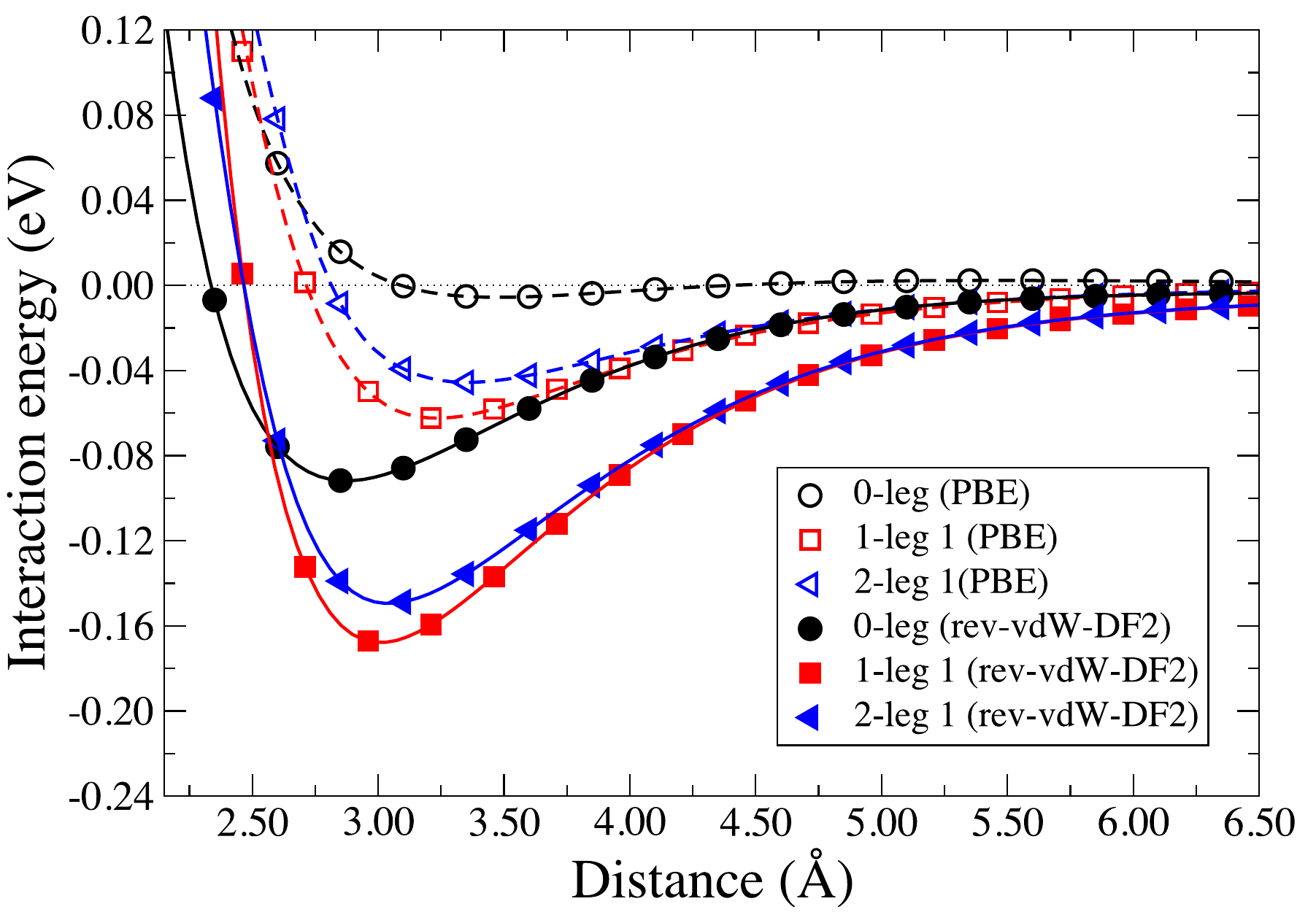}
    \caption{\label{fig:eint-pyri-n} Interaction energy of water with  graphene doped with pyridinic N and H-terminated C, as a function of water-surface distance for difference water configurations.}
 \end{figure}

Next, we investigated the adsorption of water in the 0- 1-, and 2-leg configurations on graphene doped with pyri-N (no H) and pyri-N (Fig.~\ref{fig:pyri-n}).
%
%
Shown in the main text are the most stable water configurations, and the results for other water configurations and orientations, as well as those on graphene with pyri-N (no H), are included in Supplemental Material \cite{SM}.
We focused on the center of the defect site as the adsorption site for the pyri-N cases.
%
%
The calculated interaction energy as a function of distance between water and graphene structure is shown in Fig.~\ref{fig:eint-pyri-n}, and the equilibrium interaction energy and distance are summarized in Table~\ref{tab:pyri-n-eint}.

As in the case of graphene doped with gr-N, the magnitude of the interaction energies obtained using rev-vdW-DF2 is much larger than those obtained using PBE.
The relative stabilities of water configurations using these functionals are qualitatively the same, suggesting the important role of the electrostatic interaction in determining the water configuration, as in the case of the gr-N.
We can see that in general, the water molecule prefers the orientations in which H atoms point to graphene with pyri-N, and the 1-leg (2-leg) configuration is the most (second) stable one.
%
By performing full structural optimization, the water molecule in the 1-leg configuration is inclined, but the configuration is retained (Supplemental Material \cite{SM}).
%
%

\begin{figure}[h]
  \includegraphics[width=0.94\textwidth]{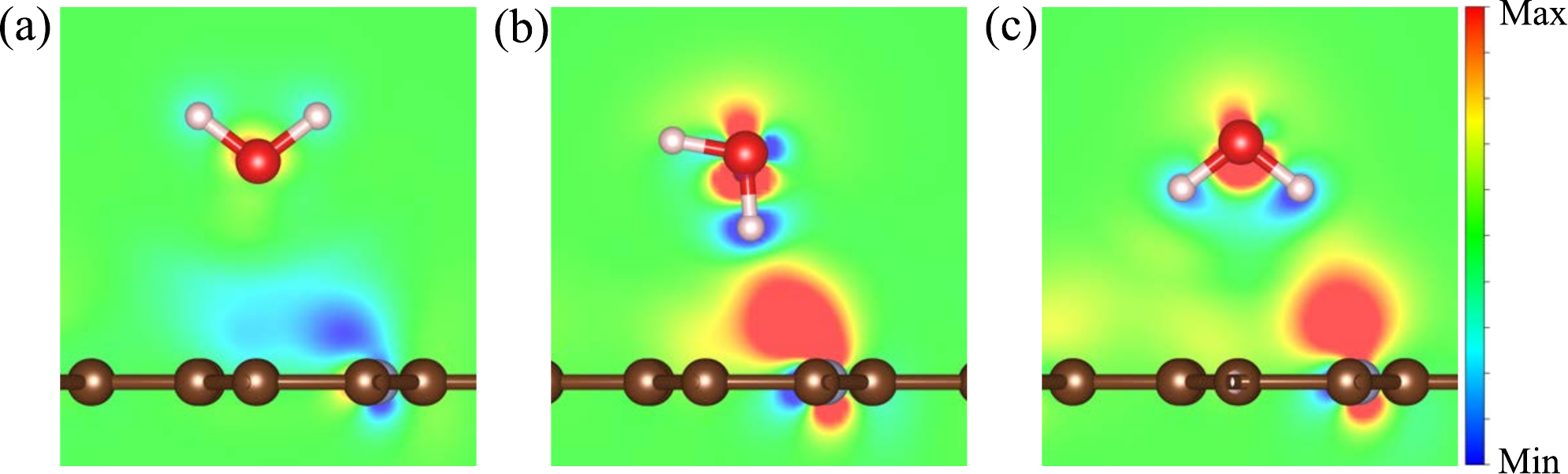}
    \caption{\label{fig:drho_pyri-n-no-h}Charge-density difference ($\Delta\rho$) of water molecule on graphene doped with pyridinic N with H-terminated C for (a) 0-leg , (b) 1-leg, (c) 2-leg configurations. $\Delta\rho$ in a plane containing the molecular plane is plotted. The maximum (minimum) value of $\Delta \rho$ is $1.35\times10^{-2}$ ($-1.35\times10^{-2}$) $e$ \AA$^{-3}$.}
 \end{figure}

%
We calculated $\Delta \rho$ for water adsorbed graphene structures with pyri-N (Fig.~\ref{fig:drho_pyri-n-no-h}).
In the pyri-N cases, the electronic charge is depleted in the vicinity of the N atom for the 0-leg configuration, while for the 1- and 2-leg configurations, the charge accumulates in the vicinity of the N atom, and the charge is more polarizable for the configurations with H atoms pointing toward the surface.
%
%
In the 1-leg 1 and 1-leg 2 configurations, the degrees of charge polarization differ, because of the relative distance and angle, which result in the difference in the interaction energies: The former is more polarizable and thus more stable.
The results suggest that pyri-N in graphene is negatively charged, contrary to gr-N, and the H atoms terminating the dangling C atoms are positively charged.
The calculated Bader charge indicates that pyri-N is negatively charged, apparently agreeing with our conclusion based on the charge density difference and the water orientation.
However, we argue that the Bader charge cannot be the sole criterion to determine the charge state of the doped N in graphene.
%
%

%
We calculated $\Delta V_{\mathrm{H}}$, and found that the potential is lower in the vicinity of the doped N atom, supporting our claim that pyri-N is negatively charged (Fig.~\ref{fig:dvhar}).
We note that in the case of pyri-N with H-terminated C atoms, the $\Delta V_{\mathrm{H}}$ in the vicinity of H atoms is positive, which also supports that terminating H atoms are positively charged and interact repulsively with H atoms in the water molecule.

\begin{table*}
\caption{\label{tab:pyri-n-eint} Equilibrium interaction energies ($E_{0}$) and distances ($d_{0}$) of water on graphene doped with pyridinic-N and H terminated C [pyri-N (no H)] and graphene doped with pyridinic-N and H terminated C (pyri-N) for 0-leg, 1-leg , and 2-leg configurations obtained using PBE and rev-vdW-DF2.}
\begin{ruledtabular}
\begin{tabular}{ccccccc}
&\multicolumn{2}{c}{0-leg} &\multicolumn{2}{c}{1-leg}  &\multicolumn{2}{c}{2-leg}\\
 \cline{2-3} \cline{4-5} \cline{6-7}
& $E_{0}$ & $d_{0}$ & $E_{0}$ & $d_{0}$ & $E_{0}$ & $d_{0}$ \\
& (meV) & (\AA) & (meV) & (\AA) & (meV) & (\AA) \\
\hline
\multicolumn{7}{c}{pyri-N (no H)}\\
PBE         &    10 &3.779 &$-$ 109 &3.059  &$-$ 67  &3.256\\
rev-vdW-DF2 &$-$ 52 &3.123 &$-$ 220 &2.921  &$-$ 170 &3.075\\
\multicolumn{7}{c}{pyri-N}\\
PBE         &$-$ 7 &3.533  &$-$ 64  &3.198 &$-$ 47  &3.368\\
rev-vdW-DF2 &$-$ 92 &2.853 &$-$ 168 &3.006 &$-$ 150 &3.038\\
\end{tabular}
\end{ruledtabular}
\end{table*}
%
%

\section{Summary}
\label{sec:summary}
We used vdW-DF to study the interaction between water and graphene doped with gr-N and pyri-N.
In contrast to pristine graphene, where the stabilities of different water configurations do not differ much, stable water configurations differ significantly, depending on the doped N configurations.
In the case of gr-N, water with oxygen pointing toward the surface is the most stable, whereas in the case of pyri-N, water with hydrogen pointing toward the surface is stable.
Based on the analyses of the electronic charges as well as electrostatic potentials, we concluded that gr-N is positively charged, whereas pyri-N, negatively charged, and that the charge state of doped N is a determining factor of the stable water configuration.
%
Our results have great implications for the interaction between water and doped graphitic materials, such as graphene, carbon nanotubes, and also for the electrochemical reactions on graphene-based catalysts, and we anticipate that they will be useful to develop DFT-based force fields for the interfaces between water and N-doped graphitic materials, to be utilized in classical molecular dynamics and/or hybrid quantum mechanics/molecular mechanics simulations \cite{Kovalenko1999,Nishihara2017}.

\begin{acknowledgments}
This work was partly supported by Grants in Aid for Scientific Research on Innovative Areas "Hydrogenomics" (Grant No. JP18H05519) and for Scientific Research (S) (Grant No. JP20H05660)
%
%
from the Ministry of Education, Culture, Sports, Science, and Technology, Japan (MEXT).
A.F.Z.A. acknowledges financial support from MEXT.\\
The data that support the findings of this study will be openly available in Materials Cloud\cite{MaterialsCloud}
\end{acknowledgments}





\bibliography{apssamp}

\providecommand{\noopsort}[1]{}\providecommand{\singleletter}[1]{#1}%
\begin{thebibliography}{62}%
\makeatletter
\providecommand \@ifxundefined [1]{%
 \@ifx{#1\undefined}
}%
\providecommand \@ifnum [1]{%
 \ifnum #1\expandafter \@firstoftwo
 \else \expandafter \@secondoftwo
 \fi
}%
\providecommand \@ifx [1]{%
 \ifx #1\expandafter \@firstoftwo
 \else \expandafter \@secondoftwo
 \fi
}%
\providecommand \natexlab [1]{#1}%
\providecommand \enquote  [1]{``#1''}%
\providecommand \bibnamefont  [1]{#1}%
\providecommand \bibfnamefont [1]{#1}%
\providecommand \citenamefont [1]{#1}%
\providecommand \href@noop [0]{\@secondoftwo}%
\providecommand \href [0]{\begingroup \@sanitize@url \@href}%
\providecommand \@href[1]{\@@startlink{#1}\@@href}%
\providecommand \@@href[1]{\endgroup#1\@@endlink}%
\providecommand \@sanitize@url [0]{\catcode `\\12\catcode `\$12\catcode
  `\&12\catcode `\#12\catcode `\^12\catcode `\_12\catcode `\%12\relax}%
\providecommand \@@startlink[1]{}%
\providecommand \@@endlink[0]{}%
\providecommand \url  [0]{\begingroup\@sanitize@url \@url }%
\providecommand \@url [1]{\endgroup\@href {#1}{\urlprefix }}%
\providecommand \urlprefix  [0]{URL }%
\providecommand \Eprint [0]{\href }%
\providecommand \doibase [0]{https://doi.org/}%
\providecommand \selectlanguage [0]{\@gobble}%
\providecommand \bibinfo  [0]{\@secondoftwo}%
\providecommand \bibfield  [0]{\@secondoftwo}%
\providecommand \translation [1]{[#1]}%
\providecommand \BibitemOpen [0]{}%
\providecommand \bibitemStop [0]{}%
\providecommand \bibitemNoStop [0]{.\EOS\space}%
\providecommand \EOS [0]{\spacefactor3000\relax}%
\providecommand \BibitemShut  [1]{\csname bibitem#1\endcsname}%
\let\auto@bib@innerbib\@empty
\bibitem [{\citenamefont {Castro~Neto}\ \emph {et~al.}(2009)\citenamefont
  {Castro~Neto}, \citenamefont {Guinea}, \citenamefont {Peres}, \citenamefont
  {Novoselov},\ and\ \citenamefont {Geim}}]{neto2009electronic}%
  \BibitemOpen
  \bibfield  {author} {\bibinfo {author} {\bibfnamefont {A.~H.}\ \bibnamefont
  {Castro~Neto}}, \bibinfo {author} {\bibfnamefont {F.}~\bibnamefont {Guinea}},
  \bibinfo {author} {\bibfnamefont {N.~M.~R.}\ \bibnamefont {Peres}}, \bibinfo
  {author} {\bibfnamefont {K.~S.}\ \bibnamefont {Novoselov}},\ and\ \bibinfo
  {author} {\bibfnamefont {A.~K.}\ \bibnamefont {Geim}},\ }\bibfield  {title}
  {\bibinfo {title} {The electronic properties of graphene},\ }\href@noop {}
  {\bibfield  {journal} {\bibinfo  {journal} {Rev. Mod. Phys.}\ }\textbf
  {\bibinfo {volume} {81}},\ \bibinfo {pages} {109} (\bibinfo {year}
  {2009})}\BibitemShut {NoStop}%
\bibitem [{\citenamefont {Novoselov}\ \emph {et~al.}(2004)\citenamefont
  {Novoselov}, \citenamefont {Geim}, \citenamefont {Morozov}, \citenamefont
  {Jiang}, \citenamefont {Zhang}, \citenamefont {Dubonos}, \citenamefont
  {Grigorieva},\ and\ \citenamefont {Firsov}}]{novoselov2004electric}%
  \BibitemOpen
  \bibfield  {author} {\bibinfo {author} {\bibfnamefont {K.~S.}\ \bibnamefont
  {Novoselov}}, \bibinfo {author} {\bibfnamefont {A.~K.}\ \bibnamefont {Geim}},
  \bibinfo {author} {\bibfnamefont {S.~V.}\ \bibnamefont {Morozov}}, \bibinfo
  {author} {\bibfnamefont {D.-e.}\ \bibnamefont {Jiang}}, \bibinfo {author}
  {\bibfnamefont {Y.}~\bibnamefont {Zhang}}, \bibinfo {author} {\bibfnamefont
  {S.~V.}\ \bibnamefont {Dubonos}}, \bibinfo {author} {\bibfnamefont {I.~V.}\
  \bibnamefont {Grigorieva}},\ and\ \bibinfo {author} {\bibfnamefont {A.~A.}\
  \bibnamefont {Firsov}},\ }\bibfield  {title} {\bibinfo {title} {Electric
  field effect in atomically thin carbon films},\ }\href@noop {} {\bibfield
  {journal} {\bibinfo  {journal} {Science}\ }\textbf {\bibinfo {volume}
  {306}},\ \bibinfo {pages} {666} (\bibinfo {year} {2004})}\BibitemShut
  {NoStop}%
\bibitem [{\citenamefont {Novoselov}\ \emph {et~al.}(2005)\citenamefont
  {Novoselov}, \citenamefont {Geim}, \citenamefont {Morozov}, \citenamefont
  {Jiang}, \citenamefont {Katsnelson}, \citenamefont {Grigorieva},
  \citenamefont {Dubonos},\ and\ \citenamefont {Firsov}}]{Novoselov2005}%
  \BibitemOpen
  \bibfield  {author} {\bibinfo {author} {\bibfnamefont {K.~S.}\ \bibnamefont
  {Novoselov}}, \bibinfo {author} {\bibfnamefont {A.~K.}\ \bibnamefont {Geim}},
  \bibinfo {author} {\bibfnamefont {S.~V.}\ \bibnamefont {Morozov}}, \bibinfo
  {author} {\bibfnamefont {D.}~\bibnamefont {Jiang}}, \bibinfo {author}
  {\bibfnamefont {M.~I.}\ \bibnamefont {Katsnelson}}, \bibinfo {author}
  {\bibfnamefont {I.~V.}\ \bibnamefont {Grigorieva}}, \bibinfo {author}
  {\bibfnamefont {S.~V.}\ \bibnamefont {Dubonos}},\ and\ \bibinfo {author}
  {\bibfnamefont {A.~A.}\ \bibnamefont {Firsov}},\ }\bibfield  {title}
  {\bibinfo {title} {{Two-dimensional gas of massless Dirac fermions in
  graphene}},\ }\href@noop {} {\bibfield  {journal} {\bibinfo  {journal}
  {Nature}\ }\textbf {\bibinfo {volume} {438}},\ \bibinfo {pages} {197}
  (\bibinfo {year} {2005})}\BibitemShut {NoStop}%
\bibitem [{\citenamefont {Georgakilas}\ \emph {et~al.}(2016)\citenamefont
  {Georgakilas}, \citenamefont {Tiwari}, \citenamefont {Kemp}, \citenamefont
  {Perman}, \citenamefont {Bourlinos}, \citenamefont {Kim},\ and\ \citenamefont
  {Zboril}}]{georgakilas2016noncovalent}%
  \BibitemOpen
  \bibfield  {author} {\bibinfo {author} {\bibfnamefont {V.}~\bibnamefont
  {Georgakilas}}, \bibinfo {author} {\bibfnamefont {J.~N.}\ \bibnamefont
  {Tiwari}}, \bibinfo {author} {\bibfnamefont {K.~C.}\ \bibnamefont {Kemp}},
  \bibinfo {author} {\bibfnamefont {J.~A.}\ \bibnamefont {Perman}}, \bibinfo
  {author} {\bibfnamefont {A.~B.}\ \bibnamefont {Bourlinos}}, \bibinfo {author}
  {\bibfnamefont {K.~S.}\ \bibnamefont {Kim}},\ and\ \bibinfo {author}
  {\bibfnamefont {R.}~\bibnamefont {Zboril}},\ }\bibfield  {title} {\bibinfo
  {title} {Noncovalent functionalization of graphene and graphene oxide for
  energy materials, biosensing, catalytic, and biomedical applications},\
  }\href@noop {} {\bibfield  {journal} {\bibinfo  {journal} {Chem. Rev.}\
  }\textbf {\bibinfo {volume} {116}},\ \bibinfo {pages} {5464} (\bibinfo {year}
  {2016})}\BibitemShut {NoStop}%
\bibitem [{\citenamefont {Geim}\ and\ \citenamefont
  {Novoselov}(2010)}]{geim2010rise}%
  \BibitemOpen
  \bibfield  {author} {\bibinfo {author} {\bibfnamefont {A.~K.}\ \bibnamefont
  {Geim}}\ and\ \bibinfo {author} {\bibfnamefont {K.~S.}\ \bibnamefont
  {Novoselov}},\ }\bibfield  {title} {\bibinfo {title} {The rise of graphene},\
  }in\ \href@noop {} {\emph {\bibinfo {booktitle} {Nanoscience and technology:
  a collection of reviews from nature journals}}}\ (\bibinfo  {publisher}
  {World Scientific},\ \bibinfo {address} {Singapore},\ \bibinfo {year}
  {2010})\ pp.\ \bibinfo {pages} {11--19}\BibitemShut {NoStop}%
\bibitem [{\citenamefont {B{\l{}}o{\'{n}}ski}\ \emph
  {et~al.}(2017)\citenamefont {B{\l{}}o{\'{n}}ski}, \citenamefont
  {Tu{\v{c}}ek}, \citenamefont {Sofer}, \citenamefont {Maz{\'{a}}nek},
  \citenamefont {Petr}, \citenamefont {Pumera}, \citenamefont {Otyepka},\ and\
  \citenamefont {Zbo{\v{r}}il}}]{blonski2017doping}%
  \BibitemOpen
  \bibfield  {author} {\bibinfo {author} {\bibfnamefont {P.}~\bibnamefont
  {B{\l{}}o{\'{n}}ski}}, \bibinfo {author} {\bibfnamefont {J.}~\bibnamefont
  {Tu{\v{c}}ek}}, \bibinfo {author} {\bibfnamefont {Z.}~\bibnamefont {Sofer}},
  \bibinfo {author} {\bibfnamefont {V.}~\bibnamefont {Maz{\'{a}}nek}}, \bibinfo
  {author} {\bibfnamefont {M.}~\bibnamefont {Petr}}, \bibinfo {author}
  {\bibfnamefont {M.}~\bibnamefont {Pumera}}, \bibinfo {author} {\bibfnamefont
  {M.}~\bibnamefont {Otyepka}},\ and\ \bibinfo {author} {\bibfnamefont
  {R.}~\bibnamefont {Zbo{\v{r}}il}},\ }\bibfield  {title} {\bibinfo {title}
  {Doping with graphitic nitrogen triggers ferromagnetism in graphene},\
  }\href@noop {} {\bibfield  {journal} {\bibinfo  {journal} {J. Am. Chem.
  Soc.}\ }\textbf {\bibinfo {volume} {139}},\ \bibinfo {pages} {3171} (\bibinfo
  {year} {2017})}\BibitemShut {NoStop}%
\bibitem [{\citenamefont {Georgakilas}\ \emph {et~al.}(2012)\citenamefont
  {Georgakilas}, \citenamefont {Otyepka}, \citenamefont {Bourlinos},
  \citenamefont {Chandra}, \citenamefont {Kim}, \citenamefont {Kemp},
  \citenamefont {Hobza}, \citenamefont {Zboril},\ and\ \citenamefont
  {Kim}}]{georgakilas2012functionalization}%
  \BibitemOpen
  \bibfield  {author} {\bibinfo {author} {\bibfnamefont {V.}~\bibnamefont
  {Georgakilas}}, \bibinfo {author} {\bibfnamefont {M.}~\bibnamefont
  {Otyepka}}, \bibinfo {author} {\bibfnamefont {A.~B.}\ \bibnamefont
  {Bourlinos}}, \bibinfo {author} {\bibfnamefont {V.}~\bibnamefont {Chandra}},
  \bibinfo {author} {\bibfnamefont {N.}~\bibnamefont {Kim}}, \bibinfo {author}
  {\bibfnamefont {K.~C.}\ \bibnamefont {Kemp}}, \bibinfo {author}
  {\bibfnamefont {P.}~\bibnamefont {Hobza}}, \bibinfo {author} {\bibfnamefont
  {R.}~\bibnamefont {Zboril}},\ and\ \bibinfo {author} {\bibfnamefont {K.~S.}\
  \bibnamefont {Kim}},\ }\bibfield  {title} {\bibinfo {title}
  {Functionalization of graphene: covalent and non-covalent approaches,
  derivatives and applications},\ }\href@noop {} {\bibfield  {journal}
  {\bibinfo  {journal} {Chem. Rev.}\ }\textbf {\bibinfo {volume} {112}},\
  \bibinfo {pages} {6156} (\bibinfo {year} {2012})}\BibitemShut {NoStop}%
\bibitem [{\citenamefont {de~la Torre}\ \emph {et~al.}(2018)\citenamefont
  {de~la Torre}, \citenamefont {{\v{S}}vec}, \citenamefont {Hapala},
  \citenamefont {Redondo}, \citenamefont {Krej{\v{c}}{\'\i}}, \citenamefont
  {Lo}, \citenamefont {Manna}, \citenamefont {Sarmah}, \citenamefont
  {Nachtigallov{\'a}}, \citenamefont {Tu{\v{c}}ek} \emph {et~al.}}]{de2018non}%
  \BibitemOpen
  \bibfield  {author} {\bibinfo {author} {\bibfnamefont {B.}~\bibnamefont
  {de~la Torre}}, \bibinfo {author} {\bibfnamefont {M.}~\bibnamefont
  {{\v{S}}vec}}, \bibinfo {author} {\bibfnamefont {P.}~\bibnamefont {Hapala}},
  \bibinfo {author} {\bibfnamefont {J.}~\bibnamefont {Redondo}}, \bibinfo
  {author} {\bibfnamefont {O.}~\bibnamefont {Krej{\v{c}}{\'\i}}}, \bibinfo
  {author} {\bibfnamefont {R.}~\bibnamefont {Lo}}, \bibinfo {author}
  {\bibfnamefont {D.}~\bibnamefont {Manna}}, \bibinfo {author} {\bibfnamefont
  {A.}~\bibnamefont {Sarmah}}, \bibinfo {author} {\bibfnamefont
  {D.}~\bibnamefont {Nachtigallov{\'a}}}, \bibinfo {author} {\bibfnamefont
  {J.}~\bibnamefont {Tu{\v{c}}ek}}, \emph {et~al.},\ }\bibfield  {title}
  {\bibinfo {title} {Non-covalent control of spin-state in metal-organic
  complex by positioning on {N}-doped graphene},\ }\href@noop {} {\bibfield
  {journal} {\bibinfo  {journal} {Nat. Commun.}\ }\textbf {\bibinfo {volume}
  {9}},\ \bibinfo {pages} {4973} (\bibinfo {year} {2018})}\BibitemShut
  {NoStop}%
\bibitem [{\citenamefont {Guo}\ \emph {et~al.}(2016)\citenamefont {Guo},
  \citenamefont {Shibuya}, \citenamefont {Akiba}, \citenamefont {Saji},
  \citenamefont {Kondo},\ and\ \citenamefont {Nakamura}}]{guo2016active}%
  \BibitemOpen
  \bibfield  {author} {\bibinfo {author} {\bibfnamefont {D.}~\bibnamefont
  {Guo}}, \bibinfo {author} {\bibfnamefont {R.}~\bibnamefont {Shibuya}},
  \bibinfo {author} {\bibfnamefont {C.}~\bibnamefont {Akiba}}, \bibinfo
  {author} {\bibfnamefont {S.}~\bibnamefont {Saji}}, \bibinfo {author}
  {\bibfnamefont {T.}~\bibnamefont {Kondo}},\ and\ \bibinfo {author}
  {\bibfnamefont {J.}~\bibnamefont {Nakamura}},\ }\bibfield  {title} {\bibinfo
  {title} {Active sites of nitrogen-doped carbon materials for oxygen reduction
  reaction clarified using model catalysts},\ }\href@noop {} {\bibfield
  {journal} {\bibinfo  {journal} {Science}\ }\textbf {\bibinfo {volume}
  {351}},\ \bibinfo {pages} {361} (\bibinfo {year} {2016})}\BibitemShut
  {NoStop}%
\bibitem [{\citenamefont {Liang}\ \emph {et~al.}(2014)\citenamefont {Liang},
  \citenamefont {Zhuang}, \citenamefont {Br{\"u}ller}, \citenamefont {Feng},\
  and\ \citenamefont {M{\"u}llen}}]{liang2014hierarchically}%
  \BibitemOpen
  \bibfield  {author} {\bibinfo {author} {\bibfnamefont {H.-W.}\ \bibnamefont
  {Liang}}, \bibinfo {author} {\bibfnamefont {X.}~\bibnamefont {Zhuang}},
  \bibinfo {author} {\bibfnamefont {S.}~\bibnamefont {Br{\"u}ller}}, \bibinfo
  {author} {\bibfnamefont {X.}~\bibnamefont {Feng}},\ and\ \bibinfo {author}
  {\bibfnamefont {K.}~\bibnamefont {M{\"u}llen}},\ }\bibfield  {title}
  {\bibinfo {title} {Hierarchically porous carbons with optimized nitrogen
  doping as highly active electrocatalysts for oxygen reduction},\ }\href@noop
  {} {\bibfield  {journal} {\bibinfo  {journal} {Nat. Commun.}\ }\textbf
  {\bibinfo {volume} {5}},\ \bibinfo {pages} {1} (\bibinfo {year}
  {2014})}\BibitemShut {NoStop}%
\bibitem [{\citenamefont {Fitri}\ \emph {et~al.}(2017)\citenamefont {Fitri},
  \citenamefont {Loh}, \citenamefont {Puspasari},\ and\ \citenamefont
  {Mohamad}}]{fitri2017cobalt}%
  \BibitemOpen
  \bibfield  {author} {\bibinfo {author} {\bibfnamefont {A.}~\bibnamefont
  {Fitri}}, \bibinfo {author} {\bibfnamefont {K.~S.}\ \bibnamefont {Loh}},
  \bibinfo {author} {\bibfnamefont {I.}~\bibnamefont {Puspasari}},\ and\
  \bibinfo {author} {\bibfnamefont {A.~B.}\ \bibnamefont {Mohamad}},\
  }\bibfield  {title} {\bibinfo {title} {Cobalt-doped carbon xerogel with
  different initial ph values toward oxygen reduction},\ }\href@noop {}
  {\bibfield  {journal} {\bibinfo  {journal} {{AIP} Conf. Proc.}\ }\textbf
  {\bibinfo {volume} {1901}},\ \bibinfo {pages} {020027} (\bibinfo {year}
  {2017})}\BibitemShut {NoStop}%
\bibitem [{\citenamefont {Abidin}\ \emph {et~al.}(2018)\citenamefont {Abidin},
  \citenamefont {Loh}, \citenamefont {Wong}, \citenamefont {Mohamad},\ and\
  \citenamefont {Puspasari}}]{abidin2018effect}%
  \BibitemOpen
  \bibfield  {author} {\bibinfo {author} {\bibfnamefont {A.~F.~Z.}\
  \bibnamefont {Abidin}}, \bibinfo {author} {\bibfnamefont {K.~S.}\
  \bibnamefont {Loh}}, \bibinfo {author} {\bibfnamefont {W.~Y.}\ \bibnamefont
  {Wong}}, \bibinfo {author} {\bibfnamefont {A.~B.}\ \bibnamefont {Mohamad}},\
  and\ \bibinfo {author} {\bibfnamefont {I.}~\bibnamefont {Puspasari}},\
  }\bibfield  {title} {\bibinfo {title} {Effect of carbon precursor and initial
  ph on cobalt-doped carbon xerogel for oxygen reduction},\ }\href@noop {}
  {\bibfield  {journal} {\bibinfo  {journal} {Int. J. of Hydrog. Energy}\
  }\textbf {\bibinfo {volume} {43}},\ \bibinfo {pages} {11047} (\bibinfo {year}
  {2018})}\BibitemShut {NoStop}%
\bibitem [{\citenamefont {Abidin}\ \emph {et~al.}(2019)\citenamefont {Abidin},
  \citenamefont {Loh}, \citenamefont {Wong},\ and\ \citenamefont
  {Mohamad}}]{abidin2019nitrogen}%
  \BibitemOpen
  \bibfield  {author} {\bibinfo {author} {\bibfnamefont {A.~F.~Z.}\
  \bibnamefont {Abidin}}, \bibinfo {author} {\bibfnamefont {K.~S.}\
  \bibnamefont {Loh}}, \bibinfo {author} {\bibfnamefont {W.~Y.}\ \bibnamefont
  {Wong}},\ and\ \bibinfo {author} {\bibfnamefont {A.~B.}\ \bibnamefont
  {Mohamad}},\ }\bibfield  {title} {\bibinfo {title} {Nitrogen-doped carbon
  xerogels catalyst for oxygen reduction reaction: improved structural and
  catalytic activity by enhancing nitrogen species and cobalt insertion},\
  }\href@noop {} {\bibfield  {journal} {\bibinfo  {journal} {Int. J. of Hydrog.
  Energy}\ }\textbf {\bibinfo {volume} {44}},\ \bibinfo {pages} {28789}
  (\bibinfo {year} {2019})}\BibitemShut {NoStop}%
\bibitem [{\citenamefont {Lin}\ \emph {et~al.}(2015)\citenamefont {Lin},
  \citenamefont {Teng}, \citenamefont {Yeh}, \citenamefont {Koshino},
  \citenamefont {Chiu},\ and\ \citenamefont {Suenaga}}]{lin2015structural}%
  \BibitemOpen
  \bibfield  {author} {\bibinfo {author} {\bibfnamefont {Y.-C.}\ \bibnamefont
  {Lin}}, \bibinfo {author} {\bibfnamefont {P.-Y.}\ \bibnamefont {Teng}},
  \bibinfo {author} {\bibfnamefont {C.-H.}\ \bibnamefont {Yeh}}, \bibinfo
  {author} {\bibfnamefont {M.}~\bibnamefont {Koshino}}, \bibinfo {author}
  {\bibfnamefont {P.-W.}\ \bibnamefont {Chiu}},\ and\ \bibinfo {author}
  {\bibfnamefont {K.}~\bibnamefont {Suenaga}},\ }\bibfield  {title} {\bibinfo
  {title} {Structural and chemical dynamics of pyridinic-nitrogen defects in
  graphene},\ }\href@noop {} {\bibfield  {journal} {\bibinfo  {journal} {Nano
  Lett.}\ }\textbf {\bibinfo {volume} {15}},\ \bibinfo {pages} {7408} (\bibinfo
  {year} {2015})}\BibitemShut {NoStop}%
\bibitem [{\citenamefont {Deng}\ \emph {et~al.}(2016)\citenamefont {Deng},
  \citenamefont {Novoselov}, \citenamefont {Fu}, \citenamefont {Zheng},
  \citenamefont {Tian},\ and\ \citenamefont {Bao}}]{deng2016catalysis}%
  \BibitemOpen
  \bibfield  {author} {\bibinfo {author} {\bibfnamefont {D.}~\bibnamefont
  {Deng}}, \bibinfo {author} {\bibfnamefont {K.}~\bibnamefont {Novoselov}},
  \bibinfo {author} {\bibfnamefont {Q.}~\bibnamefont {Fu}}, \bibinfo {author}
  {\bibfnamefont {N.}~\bibnamefont {Zheng}}, \bibinfo {author} {\bibfnamefont
  {Z.}~\bibnamefont {Tian}},\ and\ \bibinfo {author} {\bibfnamefont
  {X.}~\bibnamefont {Bao}},\ }\bibfield  {title} {\bibinfo {title} {Catalysis
  with two-dimensional materials and their heterostructures},\ }\href@noop {}
  {\bibfield  {journal} {\bibinfo  {journal} {Nat. Nanotechnol.}\ }\textbf
  {\bibinfo {volume} {11}},\ \bibinfo {pages} {218} (\bibinfo {year}
  {2016})}\BibitemShut {NoStop}%
\bibitem [{\citenamefont {Jiao}\ \emph {et~al.}(2016)\citenamefont {Jiao},
  \citenamefont {Zheng}, \citenamefont {Davey},\ and\ \citenamefont
  {Qiao}}]{jiao2016activity}%
  \BibitemOpen
  \bibfield  {author} {\bibinfo {author} {\bibfnamefont {Y.}~\bibnamefont
  {Jiao}}, \bibinfo {author} {\bibfnamefont {Y.}~\bibnamefont {Zheng}},
  \bibinfo {author} {\bibfnamefont {K.}~\bibnamefont {Davey}},\ and\ \bibinfo
  {author} {\bibfnamefont {S.-Z.}\ \bibnamefont {Qiao}},\ }\bibfield  {title}
  {\bibinfo {title} {Activity origin and catalyst design principles for
  electrocatalytic hydrogen evolution on heteroatom-doped graphene},\
  }\href@noop {} {\bibfield  {journal} {\bibinfo  {journal} {Nat. Energy}\
  }\textbf {\bibinfo {volume} {1}},\ \bibinfo {pages} {16130} (\bibinfo {year}
  {2016})}\BibitemShut {NoStop}%
\bibitem [{\citenamefont {Wu}\ \emph {et~al.}(2016{\natexlab{a}})\citenamefont
  {Wu}, \citenamefont {Ma}, \citenamefont {Sun}, \citenamefont {Gold},
  \citenamefont {Tiwary}, \citenamefont {Kim}, \citenamefont {Zhu},
  \citenamefont {Chopra}, \citenamefont {Odeh}, \citenamefont {Vajtai} \emph
  {et~al.}}]{wu2016metal}%
  \BibitemOpen
  \bibfield  {author} {\bibinfo {author} {\bibfnamefont {J.}~\bibnamefont
  {Wu}}, \bibinfo {author} {\bibfnamefont {S.}~\bibnamefont {Ma}}, \bibinfo
  {author} {\bibfnamefont {J.}~\bibnamefont {Sun}}, \bibinfo {author}
  {\bibfnamefont {J.~I.}\ \bibnamefont {Gold}}, \bibinfo {author}
  {\bibfnamefont {C.}~\bibnamefont {Tiwary}}, \bibinfo {author} {\bibfnamefont
  {B.}~\bibnamefont {Kim}}, \bibinfo {author} {\bibfnamefont {L.}~\bibnamefont
  {Zhu}}, \bibinfo {author} {\bibfnamefont {N.}~\bibnamefont {Chopra}},
  \bibinfo {author} {\bibfnamefont {I.~N.}\ \bibnamefont {Odeh}}, \bibinfo
  {author} {\bibfnamefont {R.}~\bibnamefont {Vajtai}}, \emph {et~al.},\
  }\bibfield  {title} {\bibinfo {title} {A metal-free electrocatalyst for
  carbon dioxide reduction to multi-carbon hydrocarbons and oxygenates},\
  }\href@noop {} {\bibfield  {journal} {\bibinfo  {journal} {Nat. Commun.}\
  }\textbf {\bibinfo {volume} {7}},\ \bibinfo {pages} {13869} (\bibinfo {year}
  {2016}{\natexlab{a}})}\BibitemShut {NoStop}%
\bibitem [{\citenamefont {Wu}\ \emph {et~al.}(2016{\natexlab{b}})\citenamefont
  {Wu}, \citenamefont {Liu}, \citenamefont {Sharma}, \citenamefont {Yadav},
  \citenamefont {Ma}, \citenamefont {Yang}, \citenamefont {Zou}, \citenamefont
  {Zhou}, \citenamefont {Vajtai}, \citenamefont {Yakobson} \emph
  {et~al.}}]{wu2016incorporation}%
  \BibitemOpen
  \bibfield  {author} {\bibinfo {author} {\bibfnamefont {J.}~\bibnamefont
  {Wu}}, \bibinfo {author} {\bibfnamefont {M.}~\bibnamefont {Liu}}, \bibinfo
  {author} {\bibfnamefont {P.~P.}\ \bibnamefont {Sharma}}, \bibinfo {author}
  {\bibfnamefont {R.~M.}\ \bibnamefont {Yadav}}, \bibinfo {author}
  {\bibfnamefont {L.}~\bibnamefont {Ma}}, \bibinfo {author} {\bibfnamefont
  {Y.}~\bibnamefont {Yang}}, \bibinfo {author} {\bibfnamefont {X.}~\bibnamefont
  {Zou}}, \bibinfo {author} {\bibfnamefont {X.-D.}\ \bibnamefont {Zhou}},
  \bibinfo {author} {\bibfnamefont {R.}~\bibnamefont {Vajtai}}, \bibinfo
  {author} {\bibfnamefont {B.~I.}\ \bibnamefont {Yakobson}}, \emph {et~al.},\
  }\bibfield  {title} {\bibinfo {title} {Incorporation of nitrogen defects for
  efficient reduction of co2 via two-electron pathway on three-dimensional
  graphene foam},\ }\href@noop {} {\bibfield  {journal} {\bibinfo  {journal}
  {Nano Lett.}\ }\textbf {\bibinfo {volume} {16}},\ \bibinfo {pages} {466}
  (\bibinfo {year} {2016}{\natexlab{b}})}\BibitemShut {NoStop}%
\bibitem [{\citenamefont {Marshall-Roth}\ \emph {et~al.}(2020)\citenamefont
  {Marshall-Roth}, \citenamefont {Libretto}, \citenamefont {Wrobel},
  \citenamefont {Anderton}, \citenamefont {Pegis}, \citenamefont {Ricke},
  \citenamefont {Van~Voorhis}, \citenamefont {Miller},\ and\ \citenamefont
  {Surendranath}}]{marshall2020pyridinic}%
  \BibitemOpen
  \bibfield  {author} {\bibinfo {author} {\bibfnamefont {T.}~\bibnamefont
  {Marshall-Roth}}, \bibinfo {author} {\bibfnamefont {N.~J.}\ \bibnamefont
  {Libretto}}, \bibinfo {author} {\bibfnamefont {A.~T.}\ \bibnamefont
  {Wrobel}}, \bibinfo {author} {\bibfnamefont {K.~J.}\ \bibnamefont
  {Anderton}}, \bibinfo {author} {\bibfnamefont {M.~L.}\ \bibnamefont {Pegis}},
  \bibinfo {author} {\bibfnamefont {N.~D.}\ \bibnamefont {Ricke}}, \bibinfo
  {author} {\bibfnamefont {T.}~\bibnamefont {Van~Voorhis}}, \bibinfo {author}
  {\bibfnamefont {J.~T.}\ \bibnamefont {Miller}},\ and\ \bibinfo {author}
  {\bibfnamefont {Y.}~\bibnamefont {Surendranath}},\ }\bibfield  {title}
  {\bibinfo {title} {A pyridinic {Fe}{N}$_4$ macrocycle models the active sites
  in fe/n-doped carbon electrocatalysts},\ }\href@noop {} {\bibfield  {journal}
  {\bibinfo  {journal} {Nat. Commun.}\ }\textbf {\bibinfo {volume} {11}},\
  \bibinfo {pages} {5283} (\bibinfo {year} {2020})}\BibitemShut {NoStop}%
\bibitem [{\citenamefont {Li}\ \emph {et~al.}(2013)\citenamefont {Li},
  \citenamefont {Wang}, \citenamefont {Kozbial}, \citenamefont {Shenoy},
  \citenamefont {Zhou}, \citenamefont {McGinley}, \citenamefont {Ireland},
  \citenamefont {Morganstein}, \citenamefont {Kunkel}, \citenamefont {Surwade}
  \emph {et~al.}}]{li2013effect}%
  \BibitemOpen
  \bibfield  {author} {\bibinfo {author} {\bibfnamefont {Z.}~\bibnamefont
  {Li}}, \bibinfo {author} {\bibfnamefont {Y.}~\bibnamefont {Wang}}, \bibinfo
  {author} {\bibfnamefont {A.}~\bibnamefont {Kozbial}}, \bibinfo {author}
  {\bibfnamefont {G.}~\bibnamefont {Shenoy}}, \bibinfo {author} {\bibfnamefont
  {F.}~\bibnamefont {Zhou}}, \bibinfo {author} {\bibfnamefont {R.}~\bibnamefont
  {McGinley}}, \bibinfo {author} {\bibfnamefont {P.}~\bibnamefont {Ireland}},
  \bibinfo {author} {\bibfnamefont {B.}~\bibnamefont {Morganstein}}, \bibinfo
  {author} {\bibfnamefont {A.}~\bibnamefont {Kunkel}}, \bibinfo {author}
  {\bibfnamefont {S.~P.}\ \bibnamefont {Surwade}}, \emph {et~al.},\ }\bibfield
  {title} {\bibinfo {title} {Effect of airborne contaminants on the wettability
  of supported graphene and graphite},\ }\href@noop {} {\bibfield  {journal}
  {\bibinfo  {journal} {Nat. Mater.}\ }\textbf {\bibinfo {volume} {12}},\
  \bibinfo {pages} {925} (\bibinfo {year} {2013})}\BibitemShut {NoStop}%
\bibitem [{\citenamefont {Kim}\ \emph {et~al.}(2011)\citenamefont {Kim},
  \citenamefont {Lee}, \citenamefont {Lee}, \citenamefont {Lee}, \citenamefont
  {Jang}, \citenamefont {Ahn}, \citenamefont {Kim},\ and\ \citenamefont
  {Lee}}]{kim2011chemical}%
  \BibitemOpen
  \bibfield  {author} {\bibinfo {author} {\bibfnamefont {K.-S.}\ \bibnamefont
  {Kim}}, \bibinfo {author} {\bibfnamefont {H.-J.}\ \bibnamefont {Lee}},
  \bibinfo {author} {\bibfnamefont {C.}~\bibnamefont {Lee}}, \bibinfo {author}
  {\bibfnamefont {S.-K.}\ \bibnamefont {Lee}}, \bibinfo {author} {\bibfnamefont
  {H.}~\bibnamefont {Jang}}, \bibinfo {author} {\bibfnamefont {J.-H.}\
  \bibnamefont {Ahn}}, \bibinfo {author} {\bibfnamefont {J.-H.}\ \bibnamefont
  {Kim}},\ and\ \bibinfo {author} {\bibfnamefont {H.-J.}\ \bibnamefont {Lee}},\
  }\bibfield  {title} {\bibinfo {title} {Chemical vapor deposition-grown
  graphene: the thinnest solid lubricant},\ }\href@noop {} {\bibfield
  {journal} {\bibinfo  {journal} {ACS nano}\ }\textbf {\bibinfo {volume} {5}},\
  \bibinfo {pages} {5107} (\bibinfo {year} {2011})}\BibitemShut {NoStop}%
\bibitem [{\citenamefont {Shin}\ \emph {et~al.}(2010)\citenamefont {Shin},
  \citenamefont {Wang}, \citenamefont {Huang}, \citenamefont {Kalon},
  \citenamefont {Wee}, \citenamefont {Shen}, \citenamefont {Bhatia},\ and\
  \citenamefont {Yang}}]{shin2010surface}%
  \BibitemOpen
  \bibfield  {author} {\bibinfo {author} {\bibfnamefont {Y.~J.}\ \bibnamefont
  {Shin}}, \bibinfo {author} {\bibfnamefont {Y.}~\bibnamefont {Wang}}, \bibinfo
  {author} {\bibfnamefont {H.}~\bibnamefont {Huang}}, \bibinfo {author}
  {\bibfnamefont {G.}~\bibnamefont {Kalon}}, \bibinfo {author} {\bibfnamefont
  {A.~T.~S.}\ \bibnamefont {Wee}}, \bibinfo {author} {\bibfnamefont
  {Z.}~\bibnamefont {Shen}}, \bibinfo {author} {\bibfnamefont {C.~S.}\
  \bibnamefont {Bhatia}},\ and\ \bibinfo {author} {\bibfnamefont
  {H.}~\bibnamefont {Yang}},\ }\bibfield  {title} {\bibinfo {title}
  {Surface-energy engineering of graphene},\ }\href@noop {} {\bibfield
  {journal} {\bibinfo  {journal} {Langmuir}\ }\textbf {\bibinfo {volume}
  {26}},\ \bibinfo {pages} {3798} (\bibinfo {year} {2010})}\BibitemShut
  {NoStop}%
\bibitem [{\citenamefont {Kozbial}\ \emph {et~al.}(2014)\citenamefont
  {Kozbial}, \citenamefont {Li}, \citenamefont {Sun}, \citenamefont {Gong},
  \citenamefont {Zhou}, \citenamefont {Wang}, \citenamefont {Xu}, \citenamefont
  {Liu},\ and\ \citenamefont {Li}}]{kozbial2014understanding}%
  \BibitemOpen
  \bibfield  {author} {\bibinfo {author} {\bibfnamefont {A.}~\bibnamefont
  {Kozbial}}, \bibinfo {author} {\bibfnamefont {Z.}~\bibnamefont {Li}},
  \bibinfo {author} {\bibfnamefont {J.}~\bibnamefont {Sun}}, \bibinfo {author}
  {\bibfnamefont {X.}~\bibnamefont {Gong}}, \bibinfo {author} {\bibfnamefont
  {F.}~\bibnamefont {Zhou}}, \bibinfo {author} {\bibfnamefont {Y.}~\bibnamefont
  {Wang}}, \bibinfo {author} {\bibfnamefont {H.}~\bibnamefont {Xu}}, \bibinfo
  {author} {\bibfnamefont {H.}~\bibnamefont {Liu}},\ and\ \bibinfo {author}
  {\bibfnamefont {L.}~\bibnamefont {Li}},\ }\bibfield  {title} {\bibinfo
  {title} {Understanding the intrinsic water wettability of graphite},\
  }\href@noop {} {\bibfield  {journal} {\bibinfo  {journal} {Carbon}\ }\textbf
  {\bibinfo {volume} {74}},\ \bibinfo {pages} {218} (\bibinfo {year}
  {2014})}\BibitemShut {NoStop}%
\bibitem [{\citenamefont {Munz}\ \emph {et~al.}(2015)\citenamefont {Munz},
  \citenamefont {Giusca}, \citenamefont {Myers-Ward}, \citenamefont {Gaskill},\
  and\ \citenamefont {Kazakova}}]{munz2015thickness}%
  \BibitemOpen
  \bibfield  {author} {\bibinfo {author} {\bibfnamefont {M.}~\bibnamefont
  {Munz}}, \bibinfo {author} {\bibfnamefont {C.~E.}\ \bibnamefont {Giusca}},
  \bibinfo {author} {\bibfnamefont {R.~L.}\ \bibnamefont {Myers-Ward}},
  \bibinfo {author} {\bibfnamefont {D.~K.}\ \bibnamefont {Gaskill}},\ and\
  \bibinfo {author} {\bibfnamefont {O.}~\bibnamefont {Kazakova}},\ }\bibfield
  {title} {\bibinfo {title} {Thickness-dependent hydrophobicity of epitaxial
  graphene},\ }\href@noop {} {\bibfield  {journal} {\bibinfo  {journal} {ACS
  Nano}\ }\textbf {\bibinfo {volume} {9}},\ \bibinfo {pages} {8401} (\bibinfo
  {year} {2015})}\BibitemShut {NoStop}%
\bibitem [{\citenamefont {Ho}\ and\ \citenamefont
  {Striolo}(2013)}]{ho2013polarizability}%
  \BibitemOpen
  \bibfield  {author} {\bibinfo {author} {\bibfnamefont {T.~A.}\ \bibnamefont
  {Ho}}\ and\ \bibinfo {author} {\bibfnamefont {A.}~\bibnamefont {Striolo}},\
  }\bibfield  {title} {\bibinfo {title} {Polarizability effects in molecular
  dynamics simulations of the graphene-water interface},\ }\href@noop {}
  {\bibfield  {journal} {\bibinfo  {journal} {J. Chem. Phys.}\ }\textbf
  {\bibinfo {volume} {138}},\ \bibinfo {pages} {054117} (\bibinfo {year}
  {2013})}\BibitemShut {NoStop}%
\bibitem [{\citenamefont {Akaishi}\ \emph {et~al.}(2017)\citenamefont
  {Akaishi}, \citenamefont {Yonemaru},\ and\ \citenamefont
  {Nakamura}}]{akaishi2017formation}%
  \BibitemOpen
  \bibfield  {author} {\bibinfo {author} {\bibfnamefont {A.}~\bibnamefont
  {Akaishi}}, \bibinfo {author} {\bibfnamefont {T.}~\bibnamefont {Yonemaru}},\
  and\ \bibinfo {author} {\bibfnamefont {J.}~\bibnamefont {Nakamura}},\
  }\bibfield  {title} {\bibinfo {title} {Formation of water layers on graphene
  surfaces},\ }\href@noop {} {\bibfield  {journal} {\bibinfo  {journal} {ACS
  omega}\ }\textbf {\bibinfo {volume} {2}},\ \bibinfo {pages} {2184} (\bibinfo
  {year} {2017})}\BibitemShut {NoStop}%
\bibitem [{\citenamefont {Hong}\ \emph {et~al.}(2016)\citenamefont {Hong},
  \citenamefont {Han}, \citenamefont {Schutzius}, \citenamefont {Wang},
  \citenamefont {Pan}, \citenamefont {Hu}, \citenamefont {Jie}, \citenamefont
  {Sharma}, \citenamefont {M{\"{u}}ller},\ and\ \citenamefont
  {Poulikakos}}]{hong2016mechanism}%
  \BibitemOpen
  \bibfield  {author} {\bibinfo {author} {\bibfnamefont {G.}~\bibnamefont
  {Hong}}, \bibinfo {author} {\bibfnamefont {Y.}~\bibnamefont {Han}}, \bibinfo
  {author} {\bibfnamefont {T.~M.}\ \bibnamefont {Schutzius}}, \bibinfo {author}
  {\bibfnamefont {Y.}~\bibnamefont {Wang}}, \bibinfo {author} {\bibfnamefont
  {Y.}~\bibnamefont {Pan}}, \bibinfo {author} {\bibfnamefont {M.}~\bibnamefont
  {Hu}}, \bibinfo {author} {\bibfnamefont {J.}~\bibnamefont {Jie}}, \bibinfo
  {author} {\bibfnamefont {C.~S.}\ \bibnamefont {Sharma}}, \bibinfo {author}
  {\bibfnamefont {U.}~\bibnamefont {M{\"{u}}ller}},\ and\ \bibinfo {author}
  {\bibfnamefont {D.}~\bibnamefont {Poulikakos}},\ }\bibfield  {title}
  {\bibinfo {title} {On the mechanism of hydrophilicity of graphene},\
  }\href@noop {} {\bibfield  {journal} {\bibinfo  {journal} {Nano Lett.}\
  }\textbf {\bibinfo {volume} {16}},\ \bibinfo {pages} {4447} (\bibinfo {year}
  {2016})}\BibitemShut {NoStop}%
\bibitem [{\citenamefont {Shih}\ \emph {et~al.}(2012)\citenamefont {Shih},
  \citenamefont {Wang}, \citenamefont {Lin}, \citenamefont {Park},
  \citenamefont {Jin}, \citenamefont {Strano},\ and\ \citenamefont
  {Blankschtein}}]{shih2012breakdown}%
  \BibitemOpen
  \bibfield  {author} {\bibinfo {author} {\bibfnamefont {C.-J.}\ \bibnamefont
  {Shih}}, \bibinfo {author} {\bibfnamefont {Q.~H.}\ \bibnamefont {Wang}},
  \bibinfo {author} {\bibfnamefont {S.}~\bibnamefont {Lin}}, \bibinfo {author}
  {\bibfnamefont {K.-C.}\ \bibnamefont {Park}}, \bibinfo {author}
  {\bibfnamefont {Z.}~\bibnamefont {Jin}}, \bibinfo {author} {\bibfnamefont
  {M.~S.}\ \bibnamefont {Strano}},\ and\ \bibinfo {author} {\bibfnamefont
  {D.}~\bibnamefont {Blankschtein}},\ }\bibfield  {title} {\bibinfo {title}
  {Breakdown in the wetting transparency of graphene},\ }\href@noop {}
  {\bibfield  {journal} {\bibinfo  {journal} {Phys. Rev. Lett.}\ }\textbf
  {\bibinfo {volume} {109}},\ \bibinfo {pages} {176101} (\bibinfo {year}
  {2012})}\BibitemShut {NoStop}%
\bibitem [{\citenamefont {Rafiee}\ \emph {et~al.}(2010)\citenamefont {Rafiee},
  \citenamefont {Rafiee}, \citenamefont {Yu},\ and\ \citenamefont
  {Koratkar}}]{rafiee2010superhydrophobic}%
  \BibitemOpen
  \bibfield  {author} {\bibinfo {author} {\bibfnamefont {J.}~\bibnamefont
  {Rafiee}}, \bibinfo {author} {\bibfnamefont {M.~A.}\ \bibnamefont {Rafiee}},
  \bibinfo {author} {\bibfnamefont {Z.-Z.}\ \bibnamefont {Yu}},\ and\ \bibinfo
  {author} {\bibfnamefont {N.}~\bibnamefont {Koratkar}},\ }\bibfield  {title}
  {\bibinfo {title} {Superhydrophobic to superhydrophilic wetting control in
  graphene films},\ }\href@noop {} {\bibfield  {journal} {\bibinfo  {journal}
  {Adv. Mater.}\ }\textbf {\bibinfo {volume} {22}},\ \bibinfo {pages} {2151}
  (\bibinfo {year} {2010})}\BibitemShut {NoStop}%
\bibitem [{\citenamefont {Singh}\ \emph {et~al.}(2019)\citenamefont {Singh},
  \citenamefont {Takeyasu},\ and\ \citenamefont {Nakamura}}]{Singh2019}%
  \BibitemOpen
  \bibfield  {author} {\bibinfo {author} {\bibfnamefont {S.~K.}\ \bibnamefont
  {Singh}}, \bibinfo {author} {\bibfnamefont {K.}~\bibnamefont {Takeyasu}},\
  and\ \bibinfo {author} {\bibfnamefont {J.}~\bibnamefont {Nakamura}},\
  }\bibfield  {title} {\bibinfo {title} {Active sites and mechanism of oxygen
  reduction reaction electrocatalysis on nitrogen-doped carbon materials},\
  }\href {https://doi.org/https://doi.org/10.1002/adma.201804297} {\bibfield
  {journal} {\bibinfo  {journal} {Adv. Mater.}\ }\textbf {\bibinfo {volume}
  {31}},\ \bibinfo {pages} {1804297} (\bibinfo {year} {2019})}\BibitemShut
  {NoStop}%
\bibitem [{\citenamefont {Takeyasu}\ \emph {et~al.}(2021)\citenamefont
  {Takeyasu}, \citenamefont {Furukawa}, \citenamefont {Shimoyama},
  \citenamefont {Singh},\ and\ \citenamefont {Nakamura}}]{Takeyasu2021}%
  \BibitemOpen
  \bibfield  {author} {\bibinfo {author} {\bibfnamefont {K.}~\bibnamefont
  {Takeyasu}}, \bibinfo {author} {\bibfnamefont {M.}~\bibnamefont {Furukawa}},
  \bibinfo {author} {\bibfnamefont {Y.}~\bibnamefont {Shimoyama}}, \bibinfo
  {author} {\bibfnamefont {S.~K.}\ \bibnamefont {Singh}},\ and\ \bibinfo
  {author} {\bibfnamefont {J.}~\bibnamefont {Nakamura}},\ }\bibfield  {title}
  {\bibinfo {title} {Role of pyridinic nitrogen in the mechanism of the oxygen
  reduction reaction on carbon electrocatalysts},\ }\href
  {https://doi.org/https://doi.org/10.1002/anie.202014323} {\bibfield
  {journal} {\bibinfo  {journal} {Angew. Chem. Int. Ed.}\ }\textbf {\bibinfo
  {volume} {60}},\ \bibinfo {pages} {5121} (\bibinfo {year}
  {2021})}\BibitemShut {NoStop}%
\bibitem [{\citenamefont {Hermann}\ \emph {et~al.}(2017)\citenamefont
  {Hermann}, \citenamefont {DiStasio~Jr},\ and\ \citenamefont
  {Tkatchenko}}]{Hermann2017}%
  \BibitemOpen
  \bibfield  {author} {\bibinfo {author} {\bibfnamefont {J.}~\bibnamefont
  {Hermann}}, \bibinfo {author} {\bibfnamefont {R.~A.}\ \bibnamefont
  {DiStasio~Jr}},\ and\ \bibinfo {author} {\bibfnamefont {A.}~\bibnamefont
  {Tkatchenko}},\ }\bibfield  {title} {\bibinfo {title} {First-principles
  models for van der {W}aals interactions in molecules and materials: Concepts,
  theory, and applications},\ }\href@noop {} {\bibfield  {journal} {\bibinfo
  {journal} {Chem. Rev.}\ }\textbf {\bibinfo {volume} {117}},\ \bibinfo {pages}
  {4714} (\bibinfo {year} {2017})}\BibitemShut {NoStop}%
\bibitem [{\citenamefont {Berland}\ \emph {et~al.}(2015)\citenamefont
  {Berland}, \citenamefont {Cooper}, \citenamefont {Lee}, \citenamefont
  {Schr{\" o}der}, \citenamefont {Thonhauser}, \citenamefont {Hyldgaard},\ and\
  \citenamefont {Lundqvist}}]{Berland2015}%
  \BibitemOpen
  \bibfield  {author} {\bibinfo {author} {\bibfnamefont {K.}~\bibnamefont
  {Berland}}, \bibinfo {author} {\bibfnamefont {V.~R.}\ \bibnamefont {Cooper}},
  \bibinfo {author} {\bibfnamefont {K.}~\bibnamefont {Lee}}, \bibinfo {author}
  {\bibfnamefont {E.}~\bibnamefont {Schr{\" o}der}}, \bibinfo {author}
  {\bibfnamefont {T.}~\bibnamefont {Thonhauser}}, \bibinfo {author}
  {\bibfnamefont {P.}~\bibnamefont {Hyldgaard}},\ and\ \bibinfo {author}
  {\bibfnamefont {B.~I.}\ \bibnamefont {Lundqvist}},\ }\bibfield  {title}
  {\bibinfo {title} {van der {W}aals forces in density functional theory: a
  review of the {vdW}-{DF} method},\ }\href
  {https://doi.org/10.1088/0034-4885/78/6/066501} {\bibfield  {journal}
  {\bibinfo  {journal} {Rep. Prog. Phys.}\ }\textbf {\bibinfo {volume} {78}},\
  \bibinfo {pages} {066501} (\bibinfo {year} {2015})}\BibitemShut {NoStop}%
\bibitem [{\citenamefont {Brandenburg}\ \emph
  {et~al.}(2019{\natexlab{a}})\citenamefont {Brandenburg}, \citenamefont {Zen},
  \citenamefont {Alf{\`e}},\ and\ \citenamefont
  {Michaelides}}]{brandenburg2019interaction}%
  \BibitemOpen
  \bibfield  {author} {\bibinfo {author} {\bibfnamefont {J.~G.}\ \bibnamefont
  {Brandenburg}}, \bibinfo {author} {\bibfnamefont {A.}~\bibnamefont {Zen}},
  \bibinfo {author} {\bibfnamefont {D.}~\bibnamefont {Alf{\`e}}},\ and\
  \bibinfo {author} {\bibfnamefont {A.}~\bibnamefont {Michaelides}},\
  }\bibfield  {title} {\bibinfo {title} {Interaction between water and carbon
  nanostructures: How good are current density functional approximations?},\
  }\href@noop {} {\bibfield  {journal} {\bibinfo  {journal} {J. Chem. Phys.}\
  }\textbf {\bibinfo {volume} {151}},\ \bibinfo {pages} {164702} (\bibinfo
  {year} {2019}{\natexlab{a}})}\BibitemShut {NoStop}%
\bibitem [{\citenamefont {Bl{\"o}chl}(1994)}]{blochl1994projector}%
  \BibitemOpen
  \bibfield  {author} {\bibinfo {author} {\bibfnamefont {P.~E.}\ \bibnamefont
  {Bl{\"o}chl}},\ }\bibfield  {title} {\bibinfo {title} {Projector
  augmented-wave method},\ }\href@noop {} {\bibfield  {journal} {\bibinfo
  {journal} {Phys. Rev. B}\ }\textbf {\bibinfo {volume} {50}},\ \bibinfo
  {pages} {17953} (\bibinfo {year} {1994})}\BibitemShut {NoStop}%
\bibitem [{\citenamefont {Giannozzi}\ \emph {et~al.}(2009)\citenamefont
  {Giannozzi}, \citenamefont {Baroni}, \citenamefont {Bonini}, \citenamefont
  {Calandra}, \citenamefont {Car}, \citenamefont {Cavazzoni}, \citenamefont
  {Ceresoli}, \citenamefont {Chiarotti}, \citenamefont {Cococcioni},
  \citenamefont {Dabo} \emph {et~al.}}]{giannozzi2009quantum}%
  \BibitemOpen
  \bibfield  {author} {\bibinfo {author} {\bibfnamefont {P.}~\bibnamefont
  {Giannozzi}}, \bibinfo {author} {\bibfnamefont {S.}~\bibnamefont {Baroni}},
  \bibinfo {author} {\bibfnamefont {N.}~\bibnamefont {Bonini}}, \bibinfo
  {author} {\bibfnamefont {M.}~\bibnamefont {Calandra}}, \bibinfo {author}
  {\bibfnamefont {R.}~\bibnamefont {Car}}, \bibinfo {author} {\bibfnamefont
  {C.}~\bibnamefont {Cavazzoni}}, \bibinfo {author} {\bibfnamefont
  {D.}~\bibnamefont {Ceresoli}}, \bibinfo {author} {\bibfnamefont {G.~L.}\
  \bibnamefont {Chiarotti}}, \bibinfo {author} {\bibfnamefont {M.}~\bibnamefont
  {Cococcioni}}, \bibinfo {author} {\bibfnamefont {I.}~\bibnamefont {Dabo}},
  \emph {et~al.},\ }\bibfield  {title} {\bibinfo {title} {{QUANTUM ESPRESSO}: a
  modular and open-source software project for quantum simulations of
  materials},\ }\href@noop {} {\bibfield  {journal} {\bibinfo  {journal} {J.
  Phys.: Condens. Matter}\ }\textbf {\bibinfo {volume} {21}},\ \bibinfo {pages}
  {395502} (\bibinfo {year} {2009})}\BibitemShut {NoStop}%
\bibitem [{\citenamefont {Dal~Corso}(2014)}]{dal2014pseudopotentials}%
  \BibitemOpen
  \bibfield  {author} {\bibinfo {author} {\bibfnamefont {A.}~\bibnamefont
  {Dal~Corso}},\ }\bibfield  {title} {\bibinfo {title} {{P}seudopotentials
  periodic table: {F}rom {H} to {Pu}},\ }\href@noop {} {\bibfield  {journal}
  {\bibinfo  {journal} {Comput. Mater. Sci.}\ }\textbf {\bibinfo {volume}
  {95}},\ \bibinfo {pages} {337} (\bibinfo {year} {2014})}\BibitemShut
  {NoStop}%
\bibitem [{\citenamefont {Hamada}(2014)}]{hamada2014van}%
  \BibitemOpen
  \bibfield  {author} {\bibinfo {author} {\bibfnamefont {I.}~\bibnamefont
  {Hamada}},\ }\bibfield  {title} {\bibinfo {title} {van der {W}aals density
  functional made accurate},\ }\href@noop {} {\bibfield  {journal} {\bibinfo
  {journal} {Phys. Rev. B}\ }\textbf {\bibinfo {volume} {89}},\ \bibinfo
  {pages} {121103(R)} (\bibinfo {year} {2014})}\BibitemShut {NoStop}%
\bibitem [{\citenamefont {Dion}\ \emph {et~al.}(2004)\citenamefont {Dion},
  \citenamefont {Rydberg}, \citenamefont {Schr\"oder}, \citenamefont
  {Langreth},\ and\ \citenamefont {Lundqvist}}]{Dion2004}%
  \BibitemOpen
  \bibfield  {author} {\bibinfo {author} {\bibfnamefont {M.}~\bibnamefont
  {Dion}}, \bibinfo {author} {\bibfnamefont {H.}~\bibnamefont {Rydberg}},
  \bibinfo {author} {\bibfnamefont {E.}~\bibnamefont {Schr\"oder}}, \bibinfo
  {author} {\bibfnamefont {D.~C.}\ \bibnamefont {Langreth}},\ and\ \bibinfo
  {author} {\bibfnamefont {B.~I.}\ \bibnamefont {Lundqvist}},\ }\bibfield
  {title} {\bibinfo {title} {Van der {W}aals density functional for general
  geometries},\ }\href {https://doi.org/10.1103/PhysRevLett.92.246401}
  {\bibfield  {journal} {\bibinfo  {journal} {Phys. Rev. Lett.}\ }\textbf
  {\bibinfo {volume} {92}},\ \bibinfo {pages} {246401} (\bibinfo {year}
  {2004})}\BibitemShut {NoStop}%
\bibitem [{\citenamefont {Thonhauser}\ \emph {et~al.}(2007)\citenamefont
  {Thonhauser}, \citenamefont {Cooper}, \citenamefont {Li}, \citenamefont
  {Puzder}, \citenamefont {Hyldgaard},\ and\ \citenamefont
  {Langreth}}]{Thonhauser2007}%
  \BibitemOpen
  \bibfield  {author} {\bibinfo {author} {\bibfnamefont {T.}~\bibnamefont
  {Thonhauser}}, \bibinfo {author} {\bibfnamefont {V.~R.}\ \bibnamefont
  {Cooper}}, \bibinfo {author} {\bibfnamefont {S.}~\bibnamefont {Li}}, \bibinfo
  {author} {\bibfnamefont {A.}~\bibnamefont {Puzder}}, \bibinfo {author}
  {\bibfnamefont {P.}~\bibnamefont {Hyldgaard}},\ and\ \bibinfo {author}
  {\bibfnamefont {D.~C.}\ \bibnamefont {Langreth}},\ }\bibfield  {title}
  {\bibinfo {title} {Van der {W}aals density functional: Self-consistent
  potential and the nature of the van der {W}aals bond},\ }\href
  {https://doi.org/10.1103/PhysRevB.76.125112} {\bibfield  {journal} {\bibinfo
  {journal} {Phys. Rev. B}\ }\textbf {\bibinfo {volume} {76}},\ \bibinfo
  {pages} {125112} (\bibinfo {year} {2007})}\BibitemShut {NoStop}%
\bibitem [{\citenamefont {Perdew}\ \emph {et~al.}(1996)\citenamefont {Perdew},
  \citenamefont {Burke},\ and\ \citenamefont
  {Ernzerhof}}]{perdew1996generalized}%
  \BibitemOpen
  \bibfield  {author} {\bibinfo {author} {\bibfnamefont {J.~P.}\ \bibnamefont
  {Perdew}}, \bibinfo {author} {\bibfnamefont {K.}~\bibnamefont {Burke}},\ and\
  \bibinfo {author} {\bibfnamefont {M.}~\bibnamefont {Ernzerhof}},\ }\bibfield
  {title} {\bibinfo {title} {Generalized gradient approximation made simple},\
  }\href@noop {} {\bibfield  {journal} {\bibinfo  {journal} {Phy. Rev. Lett.}\
  }\textbf {\bibinfo {volume} {77}},\ \bibinfo {pages} {3865} (\bibinfo {year}
  {1996})}\BibitemShut {NoStop}%
\bibitem [{\citenamefont {Marzari}\ \emph {et~al.}(1999)\citenamefont
  {Marzari}, \citenamefont {Vanderbilt}, \citenamefont {De~Vita},\ and\
  \citenamefont {Payne}}]{marzari1999thermal}%
  \BibitemOpen
  \bibfield  {author} {\bibinfo {author} {\bibfnamefont {N.}~\bibnamefont
  {Marzari}}, \bibinfo {author} {\bibfnamefont {D.}~\bibnamefont {Vanderbilt}},
  \bibinfo {author} {\bibfnamefont {A.}~\bibnamefont {De~Vita}},\ and\ \bibinfo
  {author} {\bibfnamefont {M.~C.}\ \bibnamefont {Payne}},\ }\bibfield  {title}
  {\bibinfo {title} {Thermal contraction and disordering of the {Al} (110)
  surface},\ }\href@noop {} {\bibfield  {journal} {\bibinfo  {journal} {Phys.
  Rev. Lett.}\ }\textbf {\bibinfo {volume} {82}},\ \bibinfo {pages} {3296}
  (\bibinfo {year} {1999})}\BibitemShut {NoStop}%
\bibitem [{SM()}]{SM}%
  \BibitemOpen
  \href@noop {} {}\bibinfo {note} {See Supplementary Material for the
  convergence study, water adsorption on pristine graphene at various
  adsorption sites, water adsorption on pristine graphene with the lattice
  parameter optimized with rev-vdW-DF2, full optimization of selected systems,
  water adsorption in various orientations on graphene doped with gr-N, pyri-N
  (no H) and pyri-N, densities of states, wave functions of adsorbed water
  molecule, charge density differences for most of the N-doped graphene systems
  considered in this work, the Hartree potential difference for graphene with
  pyri-N (no H), and Bader and L{\"o}wdin charge analyses, which includes
  Refs.~\cite{kurita2007energetics,leenaerts2008adsorption}.}\BibitemShut
  {Stop}%
\bibitem [{\citenamefont {Otani}\ and\ \citenamefont
  {Sugino}(2006)}]{otani2006first}%
  \BibitemOpen
  \bibfield  {author} {\bibinfo {author} {\bibfnamefont {M.}~\bibnamefont
  {Otani}}\ and\ \bibinfo {author} {\bibfnamefont {O.}~\bibnamefont {Sugino}},\
  }\bibfield  {title} {\bibinfo {title} {First-principles calculations of
  charged surfaces and interfaces: A plane-wave nonrepeated slab approach},\
  }\href@noop {} {\bibfield  {journal} {\bibinfo  {journal} {Phys. Rev. B}\
  }\textbf {\bibinfo {volume} {73}},\ \bibinfo {pages} {115407} (\bibinfo
  {year} {2006})}\BibitemShut {NoStop}%
\bibitem [{\citenamefont {Hamada}\ \emph {et~al.}(2009)\citenamefont {Hamada},
  \citenamefont {Otani}, \citenamefont {Sugino},\ and\ \citenamefont
  {Morikawa}}]{hamada2009green}%
  \BibitemOpen
  \bibfield  {author} {\bibinfo {author} {\bibfnamefont {I.}~\bibnamefont
  {Hamada}}, \bibinfo {author} {\bibfnamefont {M.}~\bibnamefont {Otani}},
  \bibinfo {author} {\bibfnamefont {O.}~\bibnamefont {Sugino}},\ and\ \bibinfo
  {author} {\bibfnamefont {Y.}~\bibnamefont {Morikawa}},\ }\bibfield  {title}
  {\bibinfo {title} {Green’s function method for elimination of the spurious
  multipole interaction in the surface/interface slab model},\ }\href@noop {}
  {\bibfield  {journal} {\bibinfo  {journal} {Phys. Rev. B}\ }\textbf {\bibinfo
  {volume} {80}},\ \bibinfo {pages} {165411} (\bibinfo {year}
  {2009})}\BibitemShut {NoStop}%
\bibitem [{\citenamefont {Hamada}(2012)}]{hamada2012adsorption}%
  \BibitemOpen
  \bibfield  {author} {\bibinfo {author} {\bibfnamefont {I.}~\bibnamefont
  {Hamada}},\ }\bibfield  {title} {\bibinfo {title} {Adsorption of water on
  graphene: {A} van der {W}aals density functional study},\ }\href@noop {}
  {\bibfield  {journal} {\bibinfo  {journal} {Phys. Rev. B}\ }\textbf {\bibinfo
  {volume} {86}},\ \bibinfo {pages} {195436} (\bibinfo {year}
  {2012})}\BibitemShut {NoStop}%
\bibitem [{\citenamefont {Berland}\ and\ \citenamefont
  {Hyldgaard}(2014)}]{Berland2014}%
  \BibitemOpen
  \bibfield  {author} {\bibinfo {author} {\bibfnamefont {K.}~\bibnamefont
  {Berland}}\ and\ \bibinfo {author} {\bibfnamefont {P.}~\bibnamefont
  {Hyldgaard}},\ }\bibfield  {title} {\bibinfo {title} {Exchange functional
  that tests the robustness of the plasmon description of the van der {W}aals
  density functional},\ }\href {https://doi.org/10.1103/PhysRevB.89.035412}
  {\bibfield  {journal} {\bibinfo  {journal} {Phys. Rev. B}\ }\textbf {\bibinfo
  {volume} {89}},\ \bibinfo {pages} {035412} (\bibinfo {year}
  {2014})}\BibitemShut {NoStop}%
\bibitem [{\citenamefont {Chakraborty}\ \emph {et~al.}(2020)\citenamefont
  {Chakraborty}, \citenamefont {Berland},\ and\ \citenamefont
  {Thonhauser}}]{Chakraborty2020}%
  \BibitemOpen
  \bibfield  {author} {\bibinfo {author} {\bibfnamefont {D.}~\bibnamefont
  {Chakraborty}}, \bibinfo {author} {\bibfnamefont {K.}~\bibnamefont
  {Berland}},\ and\ \bibinfo {author} {\bibfnamefont {T.}~\bibnamefont
  {Thonhauser}},\ }\bibfield  {title} {\bibinfo {title} {Next-generation
  nonlocal van der {W}aals density functional},\ }\href
  {https://doi.org/10.1021/acs.jctc.0c00471} {\bibfield  {journal} {\bibinfo
  {journal} {J. Chem. Theory Comput.}\ }\textbf {\bibinfo {volume} {16}},\
  \bibinfo {pages} {5893} (\bibinfo {year} {2020})}\BibitemShut {NoStop}%
\bibitem [{\citenamefont {Berland}\ \emph {et~al.}(2017)\citenamefont
  {Berland}, \citenamefont {Jiao}, \citenamefont {Lee}, \citenamefont {Rangel},
  \citenamefont {Neaton},\ and\ \citenamefont {Hyldgaard}}]{Berland2017}%
  \BibitemOpen
  \bibfield  {author} {\bibinfo {author} {\bibfnamefont {K.}~\bibnamefont
  {Berland}}, \bibinfo {author} {\bibfnamefont {Y.}~\bibnamefont {Jiao}},
  \bibinfo {author} {\bibfnamefont {J.-H.}\ \bibnamefont {Lee}}, \bibinfo
  {author} {\bibfnamefont {T.}~\bibnamefont {Rangel}}, \bibinfo {author}
  {\bibfnamefont {J.~B.}\ \bibnamefont {Neaton}},\ and\ \bibinfo {author}
  {\bibfnamefont {P.}~\bibnamefont {Hyldgaard}},\ }\bibfield  {title} {\bibinfo
  {title} {Assessment of two hybrid van der {W}aals density functionals for
  covalent and non-covalent binding of molecules},\ }\href
  {https://doi.org/10.1063/1.4986522} {\bibfield  {journal} {\bibinfo
  {journal} {The Journal of Chemical Physics}\ }\textbf {\bibinfo {volume}
  {146}},\ \bibinfo {pages} {234106} (\bibinfo {year} {2017})}\BibitemShut
  {NoStop}%
\bibitem [{\citenamefont {Jiao}\ \emph {et~al.}(2018)\citenamefont {Jiao},
  \citenamefont {Schröder},\ and\ \citenamefont {Hyldgaard}}]{Jiao2018}%
  \BibitemOpen
  \bibfield  {author} {\bibinfo {author} {\bibfnamefont {Y.}~\bibnamefont
  {Jiao}}, \bibinfo {author} {\bibfnamefont {E.}~\bibnamefont {Schröder}},\
  and\ \bibinfo {author} {\bibfnamefont {P.}~\bibnamefont {Hyldgaard}},\
  }\bibfield  {title} {\bibinfo {title} {Extent of {F}ock-exchange mixing for a
  hybrid van der {W}aals density functional?},\ }\href
  {https://doi.org/10.1063/1.5012870} {\bibfield  {journal} {\bibinfo
  {journal} {The Journal of Chemical Physics}\ }\textbf {\bibinfo {volume}
  {148}},\ \bibinfo {pages} {194115} (\bibinfo {year} {2018})}\BibitemShut
  {NoStop}%
\bibitem [{\citenamefont {Hyldgaard}\ \emph {et~al.}(2020)\citenamefont
  {Hyldgaard}, \citenamefont {Jiao},\ and\ \citenamefont
  {Shukla}}]{Hyldgaard2020}%
  \BibitemOpen
  \bibfield  {author} {\bibinfo {author} {\bibfnamefont {P.}~\bibnamefont
  {Hyldgaard}}, \bibinfo {author} {\bibfnamefont {Y.}~\bibnamefont {Jiao}},\
  and\ \bibinfo {author} {\bibfnamefont {V.}~\bibnamefont {Shukla}},\
  }\bibfield  {title} {\bibinfo {title} {Screening nature of the van der
  {W}aals density functional method: a review and analysis of the many-body
  physics foundation},\ }\href {https://doi.org/10.1088/1361-648x/ab8250}
  {\bibfield  {journal} {\bibinfo  {journal} {J. Phys.: Condens. Matter}\
  }\textbf {\bibinfo {volume} {32}},\ \bibinfo {pages} {393001} (\bibinfo
  {year} {2020})}\BibitemShut {NoStop}%
\bibitem [{\citenamefont {Brandenburg}\ \emph
  {et~al.}(2019{\natexlab{b}})\citenamefont {Brandenburg}, \citenamefont {Zen},
  \citenamefont {Fitzner}, \citenamefont {Ramberger}, \citenamefont {Kresse},
  \citenamefont {Tsatsoulis}, \citenamefont {Grüneis}, \citenamefont
  {Michaelides},\ and\ \citenamefont {Alf{\' e}}}]{Brandenburg2019}%
  \BibitemOpen
  \bibfield  {author} {\bibinfo {author} {\bibfnamefont {J.~G.}\ \bibnamefont
  {Brandenburg}}, \bibinfo {author} {\bibfnamefont {A.}~\bibnamefont {Zen}},
  \bibinfo {author} {\bibfnamefont {M.}~\bibnamefont {Fitzner}}, \bibinfo
  {author} {\bibfnamefont {B.}~\bibnamefont {Ramberger}}, \bibinfo {author}
  {\bibfnamefont {G.}~\bibnamefont {Kresse}}, \bibinfo {author} {\bibfnamefont
  {T.}~\bibnamefont {Tsatsoulis}}, \bibinfo {author} {\bibfnamefont
  {A.}~\bibnamefont {Grüneis}}, \bibinfo {author} {\bibfnamefont
  {A.}~\bibnamefont {Michaelides}},\ and\ \bibinfo {author} {\bibfnamefont
  {D.}~\bibnamefont {Alf{\' e}}},\ }\bibfield  {title} {\bibinfo {title}
  {Physisorption of water on graphene: Subchemical accuracy from many-body
  electronic structure methods},\ }\href
  {https://doi.org/10.1021/acs.jpclett.8b03679} {\bibfield  {journal} {\bibinfo
   {journal} {J. Phys. Chem. Lett.}\ }\textbf {\bibinfo {volume} {10}},\
  \bibinfo {pages} {358} (\bibinfo {year} {2019}{\natexlab{b}})}\BibitemShut
  {NoStop}%
\bibitem [{\citenamefont {Wu}\ \emph {et~al.}(2001)\citenamefont {Wu},
  \citenamefont {Vargas}, \citenamefont {Nayak}, \citenamefont {Lotrich},\ and\
  \citenamefont {Scoles}}]{Wu2001}%
  \BibitemOpen
  \bibfield  {author} {\bibinfo {author} {\bibfnamefont {X.}~\bibnamefont
  {Wu}}, \bibinfo {author} {\bibfnamefont {M.~C.}\ \bibnamefont {Vargas}},
  \bibinfo {author} {\bibfnamefont {S.}~\bibnamefont {Nayak}}, \bibinfo
  {author} {\bibfnamefont {V.}~\bibnamefont {Lotrich}},\ and\ \bibinfo {author}
  {\bibfnamefont {G.}~\bibnamefont {Scoles}},\ }\bibfield  {title} {\bibinfo
  {title} {Towards extending the applicability of density functional theory to
  weakly bound systems},\ }\href {https://doi.org/10.1063/1.1412004} {\bibfield
   {journal} {\bibinfo  {journal} {J. Chem. Phys.}\ }\textbf {\bibinfo {volume}
  {115}},\ \bibinfo {pages} {8748} (\bibinfo {year} {2001})}\BibitemShut
  {NoStop}%
\bibitem [{\citenamefont {Murray}\ \emph {et~al.}(2009)\citenamefont {Murray},
  \citenamefont {Lee},\ and\ \citenamefont {Langreth}}]{Murray2009}%
  \BibitemOpen
  \bibfield  {author} {\bibinfo {author} {\bibfnamefont {{\'{E}}.~D.}\
  \bibnamefont {Murray}}, \bibinfo {author} {\bibfnamefont {K.}~\bibnamefont
  {Lee}},\ and\ \bibinfo {author} {\bibfnamefont {D.~C.}\ \bibnamefont
  {Langreth}},\ }\bibfield  {title} {\bibinfo {title} {Investigation of
  exchange energy density functional accuracy for interacting molecules},\
  }\href {https://doi.org/10.1021/ct900365q} {\bibfield  {journal} {\bibinfo
  {journal} {J. Chem. Theory Comput}\ }\textbf {\bibinfo {volume} {5}},\
  \bibinfo {pages} {2754} (\bibinfo {year} {2009})}\BibitemShut {NoStop}%
\bibitem [{\citenamefont {Hamada}\ \emph {et~al.}(2010)\citenamefont {Hamada},
  \citenamefont {Lee},\ and\ \citenamefont {Morikawa}}]{Hamada2010}%
  \BibitemOpen
  \bibfield  {author} {\bibinfo {author} {\bibfnamefont {I.}~\bibnamefont
  {Hamada}}, \bibinfo {author} {\bibfnamefont {K.}~\bibnamefont {Lee}},\ and\
  \bibinfo {author} {\bibfnamefont {Y.}~\bibnamefont {Morikawa}},\ }\bibfield
  {title} {\bibinfo {title} {Interaction of water with a metal surface:
  Importance of van der {W}aals forces},\ }\href
  {https://doi.org/10.1103/PhysRevB.81.115452} {\bibfield  {journal} {\bibinfo
  {journal} {Phys. Rev. B}\ }\textbf {\bibinfo {volume} {81}},\ \bibinfo
  {pages} {115452} (\bibinfo {year} {2010})}\BibitemShut {NoStop}%
\bibitem [{\citenamefont {Pham}\ \emph {et~al.}(2021)\citenamefont {Pham},
  \citenamefont {Pham}, \citenamefont {Chihaia}, \citenamefont {Vu},
  \citenamefont {Trinh}, \citenamefont {Pham}, \citenamefont {Van~Thang},\ and\
  \citenamefont {Son}}]{Pham2021}%
  \BibitemOpen
  \bibfield  {author} {\bibinfo {author} {\bibfnamefont {T.~T.}\ \bibnamefont
  {Pham}}, \bibinfo {author} {\bibfnamefont {T.~N.}\ \bibnamefont {Pham}},
  \bibinfo {author} {\bibfnamefont {V.}~\bibnamefont {Chihaia}}, \bibinfo
  {author} {\bibfnamefont {Q.~A.}\ \bibnamefont {Vu}}, \bibinfo {author}
  {\bibfnamefont {T.~T.}\ \bibnamefont {Trinh}}, \bibinfo {author}
  {\bibfnamefont {T.~T.}\ \bibnamefont {Pham}}, \bibinfo {author}
  {\bibfnamefont {L.}~\bibnamefont {Van~Thang}},\ and\ \bibinfo {author}
  {\bibfnamefont {D.~N.}\ \bibnamefont {Son}},\ }\bibfield  {title} {\bibinfo
  {title} {How do the doping concentrations of {N} and {B} in graphene modify
  the water adsorption?},\ }\href {https://doi.org/10.1039/D1RA01506K}
  {\bibfield  {journal} {\bibinfo  {journal} {RSC Adv.}\ }\textbf {\bibinfo
  {volume} {11}},\ \bibinfo {pages} {19560} (\bibinfo {year}
  {2021})}\BibitemShut {NoStop}%
\bibitem [{\citenamefont {Mallada}\ \emph {et~al.}(2020)\citenamefont
  {Mallada}, \citenamefont {Edalatmanesh}, \citenamefont {Lazar}, \citenamefont
  {Redondo}, \citenamefont {Gallardo}, \citenamefont {Zbo{\v{r}}il},
  \citenamefont {Jel{\'{i}}nek}, \citenamefont {{\v{S}}vec},\ and\
  \citenamefont {de~la Torre}}]{mallada2020atomic}%
  \BibitemOpen
  \bibfield  {author} {\bibinfo {author} {\bibfnamefont {B.}~\bibnamefont
  {Mallada}}, \bibinfo {author} {\bibfnamefont {S.}~\bibnamefont
  {Edalatmanesh}}, \bibinfo {author} {\bibfnamefont {P.}~\bibnamefont {Lazar}},
  \bibinfo {author} {\bibfnamefont {J.}~\bibnamefont {Redondo}}, \bibinfo
  {author} {\bibfnamefont {A.}~\bibnamefont {Gallardo}}, \bibinfo {author}
  {\bibfnamefont {R.}~\bibnamefont {Zbo{\v{r}}il}}, \bibinfo {author}
  {\bibfnamefont {P.}~\bibnamefont {Jel{\'{i}}nek}}, \bibinfo {author}
  {\bibfnamefont {M.}~\bibnamefont {{\v{S}}vec}},\ and\ \bibinfo {author}
  {\bibfnamefont {B.}~\bibnamefont {de~la Torre}},\ }\bibfield  {title}
  {\bibinfo {title} {Atomic-scale charge distribution mapping of single
  substitutional p- and n-type dopants in graphene},\ }\href
  {https://doi.org/10.1021/acssuschemeng.9b07623} {\bibfield  {journal}
  {\bibinfo  {journal} {ACS Sustain. Chem. Eng.}\ }\textbf {\bibinfo {volume}
  {8}},\ \bibinfo {pages} {3437} (\bibinfo {year} {2020})}\BibitemShut
  {NoStop}%
\bibitem [{\citenamefont {Kovalenko}\ and\ \citenamefont
  {Hirata}(1999)}]{Kovalenko1999}%
  \BibitemOpen
  \bibfield  {author} {\bibinfo {author} {\bibfnamefont {A.}~\bibnamefont
  {Kovalenko}}\ and\ \bibinfo {author} {\bibfnamefont {F.}~\bibnamefont
  {Hirata}},\ }\bibfield  {title} {\bibinfo {title} {Self-consistent
  description of a metal–water interface by the {K}ohn–{S}ham density
  functional theory and the three-dimensional reference interaction site
  model},\ }\href {https://doi.org/10.1063/1.478883} {\bibfield  {journal}
  {\bibinfo  {journal} {J. Chem. Phys.}\ }\textbf {\bibinfo {volume} {110}},\
  \bibinfo {pages} {10095} (\bibinfo {year} {1999})}\BibitemShut {NoStop}%
\bibitem [{\citenamefont {Nishihara}\ and\ \citenamefont
  {Otani}(2017)}]{Nishihara2017}%
  \BibitemOpen
  \bibfield  {author} {\bibinfo {author} {\bibfnamefont {S.}~\bibnamefont
  {Nishihara}}\ and\ \bibinfo {author} {\bibfnamefont {M.}~\bibnamefont
  {Otani}},\ }\bibfield  {title} {\bibinfo {title} {Hybrid solvation models for
  bulk, interface, and membrane: Reference interaction site methods coupled
  with density functional theory},\ }\href
  {https://doi.org/10.1103/PhysRevB.96.115429} {\bibfield  {journal} {\bibinfo
  {journal} {Phys. Rev. B}\ }\textbf {\bibinfo {volume} {96}},\ \bibinfo
  {pages} {115429} (\bibinfo {year} {2017})}\BibitemShut {NoStop}%
\bibitem [{Mat()}]{MaterialsCloud}%
  \BibitemOpen
  \href@noop {} {}\bibinfo {note} {See
  http://www.materialscloud.org.}\BibitemShut {Stop}%
\bibitem [{\citenamefont {Kurita}\ \emph {et~al.}(2007)\citenamefont {Kurita},
  \citenamefont {Okada},\ and\ \citenamefont
  {Oshiyama}}]{kurita2007energetics}%
  \BibitemOpen
  \bibfield  {author} {\bibinfo {author} {\bibfnamefont {T.}~\bibnamefont
  {Kurita}}, \bibinfo {author} {\bibfnamefont {S.}~\bibnamefont {Okada}},\ and\
  \bibinfo {author} {\bibfnamefont {A.}~\bibnamefont {Oshiyama}},\ }\bibfield
  {title} {\bibinfo {title} {Energetics of ice nanotubes and their
  encapsulation in carbon nanotubes from density-functional theory},\
  }\href@noop {} {\bibfield  {journal} {\bibinfo  {journal} {Phy. Rev. B}\
  }\textbf {\bibinfo {volume} {75}},\ \bibinfo {pages} {205424} (\bibinfo
  {year} {2007})}\BibitemShut {NoStop}%
\bibitem [{\citenamefont {Leenaerts}\ \emph {et~al.}(2008)\citenamefont
  {Leenaerts}, \citenamefont {Partoens},\ and\ \citenamefont
  {Peeters}}]{leenaerts2008adsorption}%
  \BibitemOpen
  \bibfield  {author} {\bibinfo {author} {\bibfnamefont {O.}~\bibnamefont
  {Leenaerts}}, \bibinfo {author} {\bibfnamefont {B.}~\bibnamefont
  {Partoens}},\ and\ \bibinfo {author} {\bibfnamefont {F.}~\bibnamefont
  {Peeters}},\ }\bibfield  {title} {\bibinfo {title} {Adsorption of {H}$_{2}$
  {O}, {NH}$_{3}$, {CO}, {NO}$_{2}$, and {NO} on graphene: {A} first-principles
  study},\ }\href@noop {} {\bibfield  {journal} {\bibinfo  {journal} {Phys.
  Rev. B}\ }\textbf {\bibinfo {volume} {77}},\ \bibinfo {pages} {125416}
  (\bibinfo {year} {2008})}\BibitemShut {NoStop}%
\end{thebibliography}%

\end{document}